\documentclass[preprint,floats,tightenlines,11pt,eqsecnum,showpacs,aps]{revtex4}
\usepackage{amsmath}
\usepackage{graphicx}
\begin{document}
\def\appls{\hbox{$<$\kern-.75em\lower 1.00ex\hbox{$\sim$}}}







\title{QUANTUM STRUCTURE OF SPACETIME AND ITS ENTROPY IN A CYCLIC UNIVERSE WITH NEGATIVE CURVATURE}

\title{QUANTUM STRUCTURE OF SPACETIME AND ITS ENTROPY IN A CYCLIC UNIVERSE WITH NEGATIVE CURVATURE II:\\
TESTING THE THEORY}

\title{QUANTUM STRUCTURE OF SPACETIME AND ITS ENTROPY IN A CYCLIC UNIVERSE WITH NEGATIVE CURVATURE II:\\
DATA ANALYSIS AND RESULTS}

\author{Miloslav Svec\footnote{svec@hep.physics.mcgill.ca}}
\affiliation{Physics Department, Dawson College, Westmount, Quebec, Canada H3Z 1A4}

\date{November 20, 2018}

\begin{abstract}

In the Part I of this work we show that Friedmann equations and the thermodynamical Gibbs-Duhem relation determine a general form of the Hubble function called Model E which predicts a dynamical Dark Energy and a dynamical Dark Matter with equations of state $w_0=-1$ and $w_M=0$, respectively. We identify Dark Energy and Dark Matter with the Space. General theory of relativity asserts that the Space is gravitational fields. We propose the Space has a specific quantum strucrure: entangled Space quanta form Dark Energy, non-entangled ones form Dark Matter. We identify Dark Matter and Dark Energy as the gravitational fields generated by Fisher information metric from the probability distributions $p$ and $q$ of the entropies carried by their quanta, respectively. 

This model of the quantum structure of the spacetime determines a specific form of the dynamical terms of Dark Energy and Dark Matter and predicts the existence of a new "residual" matter term with equation of state $w_r=-\frac{1}{3}$. This term plays the role of a curvature term in the Hubble function with negative curvature $k=-1$. Its consistency with the curvature term in the Robertson-Walker metric then predicts positive present curvature density $\Omega_{c,0}$ which in turn places constraints on the cosmological parameters. In this work we test these predictions in fits to the Hubble data $H(z)$ and angular diameter distance data $d_A(z)$. The fits confirm all predictions and allow us to identify the Model E with the analytical Model A of a Cyclic Universe developed independently in an earlier work. These results support our model of the quantum structure of the spacetime and identify it with the anti-de Sitter spacetime of the Cyclic Universe.

\end{abstract}
\pacs{9880.-k, 9880.Qc, 9535.+d, 9535.+x}

\maketitle

\tableofcontents

\newpage
\section{Introduction}

The Universe consists of the Space and Matter. Einstein's General theory of  relativity asserts that Space is gravitational fields~\cite{einstein16}. In Part I of this work~\cite{svec17e} we propose to identify Dark Energy and Dark Matter with the Space. The gravitational fields of Dark Energy and Dark Matter arise from the quantum structure of the spacetime. 

In our model of the quantum spacetime~\cite{svec17e} we identify the space quanta with two-qubit quantum states of massless gravitons with helicities $|\pm2>$. All space quanta carry quantum information entropy $S$. All entangled quanta carry also entanglement entropy $S_E$ and form Dark Energy. All non-entangled quanta form Dark Matter. The average quantum state of Dark Energy is a special state $\rho_\lambda(t)$. It is described by the scale factor $a(t)$ and carries entropy $\Sigma_\lambda(t)$. 

In the absence of Baryonic matter Dark Matter and Dark Energy are described by the probability distributions $p(\vec{x},t,S)$ and $q(\vec{x},t,\chi)$, respectively. Here $S$ is the von Neumann entropy of Dark Matter states and $\chi=S+S_E$ is the combined entropy of Dark Energy states~\cite{svec17e}. Fisher information metric~\cite{svec17e} generates from these distributions the gravitational fields $h^{MV}_{\mu\nu}$ and $h^{EV}_{\mu\nu}$ of the free Dark Matter and Dark Energy. In the presence of the Baryonic matter the distributions are displaced $p\to p'$ and $q\to q'$. Fisher metric then defines the displaced fields $h^{MB}_{\mu\nu}$ and $h^{EB}_{\mu\nu}$. In general, on galactic and cluster scales we expect a homogeneous Dark Energy with $h^{EV}_{\mu \nu} \approx 0$ and $h^{EB}_{\mu \nu} \approx 0$.

The relations between the distribution functions and the gravitational fields of Dark Matter and Dark Energy given by the Fisher matric are called quantum duality relations. The quantum duality relations for the Baryonic matter relate the Baryonic gravitational field to their perturbations $\delta p$ and $\delta q$ of the free distributions $p$ and $q$, respectively. The details of this quantum theory of the Space are presented in the Part I of this work~\cite{svec17e}.

The space quanta are observable physical entities. They manifest themselves in a series of predictions for the Hubble function and are thus testable in Hubble data $H(z)$ and in luminosity distance data $d_L(z)$ and angular diameter distance data $d_A(z)$. We present such tests in this Part II of our work.

In the Part I~\cite{svec17e} we have constructed a general solution for the Hubble function from the joint dynamics of the Friedmann equations and the Euler equation of the Thermodynamics. Its general form reads
\begin{equation}
H^2 = H_0^2\Bigl[\Omega_{0,0}+\Sigma_0(z)
+(1+z)^3\Bigl(\Omega_{Mm,0}+\Sigma_{M}(z)\Bigr)+(1+z)^4\Omega_{rad,0}\Bigr]\\
\end{equation}
The theory predicts the existence of the dynamical Dark Energy $\tilde{\Omega}_0=\Omega_{0,0}+\Sigma_0(z)$ and dynamical Dark Matter 
$\tilde{\Omega}_M=\Omega_{M,0}+\Sigma_M(z)$ with equations of state $w_0=-1$ and $w_M=0$, respectively, at all $z$. The new terms $\Sigma_0(z)$ and $\Sigma_M(z)$ are so called entropic terms. The subscript $Mm$ in (1.1) indicates the inclusion of the atomic matter $\Omega_m$.

The quantum theory of the Space is testable because it modifies the Hubble function (1.1) in two ways:\\
(1) It predicts a specific form of the entropic terms $\Sigma_0(z)$ and $\Sigma_M(z)$ in terms of the entropy $\Sigma_\lambda(t)$ of the average quantum state $\rho_\lambda(t)$ of Dark Energy.\\
(2) It predicts the existence of a new "residual" matter with the 
equation of state $w_r=-\frac{1}{3}$ which adds a new term to the
Hubble function (1.1)
\begin{equation}
\tilde{\Omega}_r=(1+z)^{3(1+w_r)}\Omega_{r,0}=(1+z)^2\Omega_{r,0}>0
\end{equation}

The new term (1.2) is akin to the spatial curvature
term $(1+z)^2\Omega_{c,0}$ arising from the Robertson-Walker metric. Since Hubble function is defined exclusively by the scale factor this term does not contribute to it. We can interpret the "residual" matter term as an "internal" curvature generated by the galactic dynamics of Dark Matter~\cite{svec17e}. Since the $\Omega_{r,0}$ is positive the consistency with the "external" curvature then implies negative spatial curvature $k=-1$ of anti-de Sitter spacetime with $\Omega_{c,0}>0$. The predicted positivity of $\Omega_{c,0}$ implies two constraints on the cosmological parameters in terms of the $\Lambda$CDM Model~\cite{svec17e}
\begin{eqnarray}
\Omega_{M,0}+\Omega_{r,0} & > & \Omega_M(\Lambda CDM)\\
\Omega_{0.0} & < & \Omega_\Lambda(\Lambda CDM)
\end{eqnarray}
The fits of the Model E to the Hubble Data $H(z)$ and angular diameter distance data $d_A(z)$ are called Model E1.

In a related paper~\cite{svec17a} we develop an analytical Model A of the Cyclic Universe with the scale factor $a(z)$ determined at all redshifts $0\leq z \leq z_\alpha$ where $z_\alpha=z(t_\alpha=0)$. The fits of the Model A to the Hubble data $H(z)$ and angular diameter distance data $d_A(z)$ generate a new set of data $AH(z)$ and $Ad_A(z)$ of the fitted values. The fits of the Model E to the data $AH(z)$ and $Ad_A(z)$ are called Model E2. 

Both Models E1 and E2 confirm the existence of the "residual" matter with equation of state $w_r=-\frac{1}{3}$. Both Models E1 and E2 as well as the Model A confirm $\Omega_{c,0}>0$ and predict cosmological parameters that satisfy the constraints (1.3) and (1.4). The Model A has the overall best fits. The Model E2 agrees with the Model A at an  astonishing $\chi^2/dof=0.0000056$. This allows us to identify the Model E2 with the Model A to describe a Cyclic Universe with quantum structure of its anti-de Sitter spacetime.

The paper is organized as follows. In Section II we review briefly the Friedmann equations for a Cyclic Universe. In Section III we outline the analytical Model A of the Cyclic Universe~\cite{svec17a} and use the Friedmann equations to calculate the cosmological parameters of the model. In Section IV we present the entropic Model E and the expressions for the entropic terms $\Sigma_0(z)$ and $\Sigma_M(z)$ in terms of the entropy $\Sigma_\lambda(z)$ of the Dark Energy average state $\rho_\lambda(t)$ derived in Part I~\cite{svec17e}. These terms are absent in the non-entropic model L. 
In Section V we fit Model E and Model L to two sets of data $H(z)$ and $AH(z)$ which define the Models E1, L1 and E2, L2, respectively. In addition we fit Model A and the $\Lambda$CDM Model. We examine the fits with different values of the equation of state $w_m$ and $\Omega_{rad,0}$. In Section VI we fit the best fits from the Section V again to two sets of data $d_A(z)$ and $AdA(z)$ to determine their predicted values of the curvature density $\Omega_{c,0}$. In Section VII we use the Friedmann equations to determine the exact entropic terms $\Sigma_0$ and $\Sigma_M$ in the Model A at all z. Since the Model A and Model E2 coincide in the linear approximation of the entropic terms we identify the Model E2 with Model A at all $z$. We discuss aspects of the Cyclic Universe with anti-de Sitter spacetime in the Section VIII. In Section IX we describe time evolution of the spatial curvature density and illustrate it with Model A. The paper closes with our conclusions and the outlook in the Section X and an Appendix.

\section{Friedmann equations of the Cyclic Universe}

We assume a homogeneous and isotropic spacetime with Robertson-Walker (RW) metric. In the cartesian coordinates it is given by~\cite{weinberg08,carroll04}
\begin{eqnarray}
g_{ij} & = & a(t)^2(t)\Bigl(\delta_{ij}+k\frac{x^ix^j}
{R_0^2-k\vec{x}^2} \Bigr)\\
\nonumber
g_{i0} & = & 0, g_{00}=-1
\end{eqnarray}
where $R_0$ is the curvature parameter and $k=-1,0,+1$ stands for open, flat and closed geometry. For a homogeneous and isotropic cosmic fluid with energy density $\rho$ and pressure $p$ Friedmann equations have the form
\begin{eqnarray}
\rho+\rho_\Lambda+\rho_c & = & \frac{3c^2}{8\pi G} H^2\\
p+p_\Lambda+p_c & = & \frac{3c^2}{8\pi G}\bigl(-H^2-\frac{2}{3}\frac{dH}{dt} \bigr)
\end{eqnarray}
Here $\rho_\Lambda$ and $p_\Lambda=-\rho_\Lambda$ are the energy density and pressure of the cosmological constant. $\rho_c$ and 
$p_c=-\frac{1}{3}\rho_c$ are the energy density and pressure of the curvature~\cite{carroll04}. These two energy densities are given by
\begin{equation}
\rho_\Lambda = \frac{3c^2}{8\pi G} \frac{\Lambda}{3}, \quad
\rho_c = \frac{3c^2}{8\pi G}\frac{-kc^2}{R_0^2a^2}
\end{equation}
where $\Lambda$ is the cosmological constant. The Hubble function is defined in terms of the scale factor
\begin{equation}
H(t)=\frac{1}{a(t)}\frac{da(t)}{dt}
\end{equation}
The cyclic scale factor is a complex wave function with a period $T$ so that $a(t+T)=a(t)$. During the expansion phase $H(t)>0$, during the contraction $H(t)<0$. At the turning points $t_\alpha=0$ and $t_\omega=T/2$ of the expanding  Universe the scale factor $a(t_\alpha)=a_{min}> 0$ and $a(t_\omega)=a_{max}< \infty$. Consequently
\begin{equation}
H(t_\alpha)=H(t_\omega)=0
\end{equation}
The contraction phase ends at the turning point $t_{2\alpha}=T$ with the scale factor $a(t_{2\alpha})=a_{min}$ and $H(t_{2\alpha})=0$. 

Since $H(t)$ is a cyclic function the combinations
\begin{eqnarray}
\bar{\rho}  & = & \rho + \rho_\Lambda + \rho_c\\
\nonumber
\bar{p} & = & p + p_\Lambda + p_c
\end{eqnarray}
are the cyclic energy density and the cyclic pressure. The Friedmann equations for the Cyclic Universe then read
\begin{eqnarray}
\bar{\rho} & = & \frac{3c^2}{8\pi G} H^2\\
\bar{p} & = & \frac{3c^2}{8\pi G}\bigl(-H^2-\frac{2}{3}\frac{dH}{dt} \bigr)
\end{eqnarray}
where $\bar{\rho}$ and $\bar{p}$ satisfy the continuity equation 
\begin{equation}
\frac{d\bar{\rho}}{dt} +3H\bar{\rho} = -3H\bar{p}
\end{equation}
Notice that $H(t)$ does not depend on the curvature parameter $R_0$ and therefore on $\rho_c$ and $p_c$.

\section{Analytical Model A of the Cyclic Universe.}

\subsection{The Hubble functions $H(t)$ and $H(z)$ of the Model A}

In a recent work~\cite{svec17a} we have constructed several analytical versions of a cyclic scale factor $a(t)$ which define an explicit analytical form of the Hubble functions $H(t)$ and $H(z)$. A version called Model A has the best $\chi^2$ fit to the Hubble data $H(z)$. 
For $z \ll z_\alpha=z(t_\alpha)$ the function $H(z)$ reads
\begin{equation}
H(z)=H_0 \frac{1+Dz}{1+z} \Bigl[ \frac{4(1+Dz)^{\frac{1}{n}}-F_0}
{4-F_0} \Bigr]^{\frac{1}{2}}
\end{equation}
where $D=1-CF_0^n$. We work with the fit labeled $A.01m$ with fitted parameters $F_0$, $n$, $C$
\begin{eqnarray}
\begin{array}{rcl}
F_0=0.490581, & n=0.284589, & C=0.527679\\
a_0=1.434732, & a_\omega=6.834258, & \bar{a}_0=0.209932
\end{array}
\end{eqnarray}
where $a_0=a(t_0)$, $a_\omega=a(t_\omega)$, 
$\bar{a}(t_0)=a_0/a_\omega$ are predicted values and $t_0$ is the present time. The Hubble parameter  $H_0=67.81$ km/sMpc is from the Planck CMB measurement~\cite{planck15}.

In the Model A the scale factor $a(t)$ is given by
\begin{equation}
a(t)=\frac{F^n(t)}{1-CF^n(t)}
\end{equation}
where $F=1+p^2-2p\cos \phi$. The Hubble function is given by (2.5) and reads at all times $t$
\begin{equation}
H(t)=\Omega \sin \phi \frac {2np}{F(1-CF^n)}
\end{equation}
where $\Omega=0.0518535$ radGyr$^{-1}$ is the angular frequency and $\phi=\Omega t$ is the phase.  Parameter $p=1-\delta$ where $\delta=3.5736$x10$^{-56}$ is determined by the Planck temperature~\cite{svec17a}. Time period $T=\frac{2\pi}{\Omega}=121.1719$ Gyr is the time of one complete cycle of the expansion and contraction of the Universe.

\subsection{Energy density of Dark Energy and Dark Matter in terms of deceleration parameter}

Deceleration parameter $q(t)$ or $q(z)$ is closely related to the Hubble function $H(t)$ or $H(z)$, respectively, by the relations
\begin{eqnarray}
q(t) & = & \frac{-1}{H^2a}\frac{d^2a}{dt^2}=-1-\frac{1}{H^2}\frac{dH}{dt}\\
q(z) & = &
-1+\frac{1+z}{H}\frac{dH}{dz}=-1+\frac{1}{2}\frac{1+z}{H^2}\frac{dH^2}{dz}
\end{eqnarray}
With normalized and unnormalized fractional energy densities $\Omega_k$ and $\tilde{\Omega}_k$ defined, respectively, by
\begin{equation}
\Omega_k\equiv\frac{8 \pi G}{3 c^2 H^2} \rho_k=\frac{H_0^2}{H^2}\tilde{\Omega}_k
\end{equation}
for $k=0,M,m,rad$ the Friedmann equations (2.8) and (2.9) take the form for all $z$ (or for all $t$)
\begin{eqnarray}
\Omega_0(z)+\Omega_{M}(z)+\Omega_{m}(z)+\Omega_{rad}(z) & = & 1\\
\nonumber
w=-\Omega_0(z)+w_m\Omega_{m}(z)+\frac{1}{3}\Omega_{rad}(z) & = & -\frac{1}{3}(1-2q(z))
\end{eqnarray}
where $w$ is equation of state in $\bar{p}=w\bar{\rho}$. We have used these relations to verify the correctnes of our fitting computer programs. Solving (3.8) for $\Omega_0$ and $\Omega_M$ we have
\begin{eqnarray}
\Omega_0 & = & \frac{1}{3}\Bigl( 1-2q+3w_m \Omega_m +\Omega_{rad}\Bigr)\\
\nonumber
\Omega_M & = & \frac{2}{3}\Bigl(1+q -\frac{3}{2}(1+w_m)\Omega_m-2\Omega_{rad}\Bigl)
\end{eqnarray}
For $w_m=0$ $\Omega_{M}$ has the meaning of the pure Dark Matter. For $w_m=w_r=-\frac{1}{3}$ $\Omega_{M}$ combines Dark Matter and atomic matter and $\Omega_{m}$ has the meaning of the "residual" matter density $\Omega_{r}$.

Analytical forms of $q(t)$ for all $t$ and $q(z)$ for all $z$ as well as for $z\ll z_\alpha$ for the Model A are given and discussed in detail in~\cite{svec17a}. Given $\Omega_m$ and $\Omega_{rad}$ the equations (3.9) determine analytically for all $t$ and all $z$ Dark Energy $\Omega_0$ and Dark Matter $\Omega_M$ (in Ref.~\cite{svec17a}) and their entropic terms $\Sigma_{0}$ and $\Sigma_{M}$  (in Section VII). With the fitted value of $q_0=q(t_0)=-0.2912$ and assuming $\Omega_{r,0}=\Omega_{m,0}=0.0484$~\cite{PDG2015} and $\Omega_{rad,0}=0.0055$ the relations (3.8) predict for the Cyclic Model A
\begin{eqnarray}
\begin{array} {lcr}
\Omega_{0,0}=0.5293, & \Omega_{M,0}=0.4168, & w_m=0\\ \Omega_{0,0}=0.5132, & \Omega_{M,0}=0.4329, & w_m=-\frac{1}{3}
\end{array}
\end{eqnarray}
These values of the cosmological parameters satisfy the constraits (1.3) and (1.4), hinting at positive $\Omega_{c,0}$ in the Model A.

Similarly to (3.7) for the components $k=0,M,m,rad$ we can define fractional densities $\Omega(z)$, $\Omega_{c}(z)$ and $\Omega_{\Lambda}(z)$ corresponding to the densities $\rho(t)$, $\rho_c(t)$ and $\rho_\Lambda=const$ in (2.2). Then the energy fraction $\Omega(z)= 1-\Omega_{c}(z)-\Omega_{\Lambda}(z)$. While we can determine with our Models $\Omega_{c}(z)$ from astrophysical data we cannot do so for $\Omega_{\Lambda,0}$. The energy density of the cosmic fluid $\rho(t)$ is then determined up to the unknown constant $\Omega_{\Lambda,0}$.

\section{Quantum model of the Entropic Hubble function}

The Hubble function in the Entropic theory introduced in the Part I of this work~\cite{svec17e} reads
\begin{equation}
H^2=H_0^2\Bigl[\Omega_{0,0}+\Sigma_0(z)+(1+z)^3\Bigl(\Omega_{M,0}(z)+\Sigma_M(z)\Bigr)+(1+z)^{3(1+w_m)}\Omega_{m,0}+(1+z)^4\Omega_{rad,0}\Bigr]
\end{equation}
For $w_m=0$ the term $\Omega_M(z)$ has the meaning of the pure Dark Matter. For $w_m \neq 0$ the term $\Omega_M(z)$ means $\Omega_{Mm}(z)=\Omega_M+\Omega_m$ as it includes both Dark Matter and atomic matter. The entropic terms are related to the entropy $S_0$ of Dark Energy and $S_M$ of Dark Matter by relations~\cite{svec17e}
\begin{eqnarray}
\Sigma_0(z) & = & \frac{8\pi G}{3c^2H_0^2}\int \limits_0^z 
\frac{kT}{V}\frac{dS_0}{dz'}dz'\\
\Sigma_M(z) & = & \frac{8\pi G}{3c^2H_0^2}\int \limits_0^z 
\frac{1}{(1+z')^{3}}\frac{kT}{V}\frac{dS_M}{dz'}dz'
\end{eqnarray}
The entropy $S_0(t)$ is a functional $S_0(\Sigma_\lambda(t))$ of the entropy $\Sigma_\lambda(t)$ of the average quantum state $\rho_\lambda(t)$ of Dark Energy. The entropies $S_0(t)+S_M(t)=S'=const$ so that $\frac{dS_M}{dz}=-\frac{dS_0}{dz}$. In the Part I of this work~\cite{svec17e} we evaluate these derivatives in the linear approximation of $\Sigma_\lambda(t)$ which allows us to calculate the integrals (4.2) and (4.3) and express both entropic terms as specific functions of $z$. These expressions are a specific prediction of our quantum model of spacetime and define the quantum model of the Hubble function. 

The Entropic Models E are then defined by the expressions for $\tilde{\Omega}_0(z)$ and $\tilde{\Omega}_{M}(z)$
\begin{eqnarray}
\tilde{\Omega}_0(z) & = & \Omega_{0,0}+A_T\Sigma_{0,A}+B_T\Sigma_{0,B}=
\Omega_{0,0}+\Sigma_{0}\\
\tilde{\Omega}_{M}(z) & = & (1+z)^3\Bigl[\Omega_{M,0}+A_T\Sigma_{M,A}+B_T\Sigma_{M,B}\Bigr]=
(1+z)^3\Bigl[\Omega_{M,0}+\Sigma_{M}\Bigr]
\end{eqnarray}
where $A_T$ and $B_T$ are free parameters. The explicit forms of the entropic integrals $\Sigma_{k,A}$ and $\Sigma_{k,B}$, $k=0,M$ read~\cite{svec17e} 
\begin{eqnarray}
\Sigma_{0A} & = & -((1+z)^2-1)\frac{2-\ln2+2\ln \bar{a}_0}{2}\bar{a}_0^2
                  +(1+z)^2(\ln (1+z))\bar{a}_0^2\\
\Sigma_{0B} & = & -\frac{4}{3}(\ln (1+z))^3\bar{a}_0^4 +(\ln (1+z))^2 
                   (1-2\ln2+4\ln\bar{a}_0)\bar{a}_0^4\\
\nonumber                    
            &   & -\ln (1+z)\Bigl[2\ln\bar{a}_0\bigl(1-2\ln2+
                   2\ln\bar{a}_0 \bigr)-\ln 2(1-\ln 2)\Bigr]\bar{a}_0^4\\
\Sigma_{MA} & = & +\frac{\ln(1+z)}{1+z}2\bar{a}_0^2 
                  -\frac{z}{1+z}\bar{a}_0^2(1+\ln2-2\ln\bar{a}_0)\\
\Sigma_{MB} & = & -\frac{(\ln(1+z))^2}{(1+z)^3}\frac{4 \bar{a}_0^4}{3}
                  +\frac{\ln(1+z)}{(1+z)^3}\bar{a}_0^4
    \Bigl[\frac{8\ln\bar{a}_0}{3}-\frac{2}{9}-\frac{4 \ln 2}{3}\Bigr]\\
\nonumber
            &   & -\Bigl[\frac{1}{(1+z)^3}-1.0\Bigr]\bar{a}_0^4
     \Bigl[\frac{4}{3}(\ln \bar{a}_0)^2-\frac{2}{3} \ln(\bar{a}_0)\bigl(\frac{1}{3}+2\ln2\bigr)
     +\frac{2}{27}+\frac{\ln2}{9}+\frac{(\ln2)^2}{3}\Bigr]
\end{eqnarray}
We have verified numerically that these integrals satisfy the condition from the conservation of the entropy
\begin{equation}
\frac{d\Sigma_0}{dz}+(1+z)^3\frac{d\Sigma_M}{dz}=0
\end{equation}
The parameter $\bar{a}_0$ comes from the Model A and is given in (3.2).

The Non-Entropic Models L are defined by $A_T=B_T=0$. With the entropic terms $\Sigma_0(z)$ and $\Sigma_M(z)$ absent the Non-Entropic Hubble function has a formal structure akin to $\Lambda$CDM model. However the Dark Energy term $\Omega_{0,0}$ does not have the meaning of the cosmological constant.

\section{Hubble data analysis and the results}

The most recent measured values of the Hubble parameter used in our fits are from Ref.~\cite{jimenez03}-Ref.~\cite{moresco16} and are tabulated in the Table VI in the Appendix A in our previous work~\cite{svec17a}. Our objectives are to fit the Entropic Model E and the Non-Entropic Model L to the Hubble data as well as to the Model A to see how close these models are to the cyclic model. The details of the fit of Model A to Hubble data are given in Ref.~\cite{svec17a} where the Table VI also lists the fitted values and errors of the Hubble function of the Model A which represent the Model A. We shall refer to these values as the Hubble data $AH(z)$. The remarkable feature of this data are the very small errors. The input value of $H_0$ was fixed at $H_0=67.81$ kms$^{-1}$Mpc$^{-1}$~\cite{planck15}. We recover this value from the predictions for $H(z=0)$ from all fitted models of $H(z)$.

We shall refer to Entropic and Non-Entropic fits of $H(z)$ data as Models E1 and L1. The corresponding fits to the $AH(z)$ data are referred to as Models E2 and L2. 

The present atomic matter energy density $\Omega_{m,0}$ and the radiation energy density  $\Omega_{rad,0}=\Omega_{\gamma,0}+\Omega_{\nu,0}$ will not be fitted but will be kept at fixed best estimated values. We take $\Omega_{m,0}=0.0484$ and the photon $\Omega_{\gamma,0}=5.38\times 10^{-5}$ determined from the measurements of the Cosmic Microwave Background~\cite{PDG2015}. The latest estimate of the neutrino background is at $\Omega_{\nu,0} < 0.0160$ (CMB) and $\Omega_{\nu,0} \geq 0.0012$ (mixing)~\cite{PDG2015}. The previous estimate was at $\Omega_{\nu,0} \leq 0.0050$~\cite{PDG2014}. 

In our first study we examine the dependence of the Entropic and Non-Entropic models on select fixed input values of $\Omega_{rad,0}$ at fixed  $w_m=0$. In our second study we examine the dependence of the Models E1 and E2 on select fixed input values of the parameter $w_m$ at fixed $\Omega_{rad,0}=0.0055$ corresponding to the best fit of the Model E2 at $w_m=0$. In all our fits we have verified that the two conditions (3.9) are satisfied exactly at all values of $z$ confirming the equations of state $w_0=-1$, $w_M=0$, $w_m=input$, $w_{rad}=\frac{1}{3}$.

\begin{table} 
\caption{Models E1 and E2: Values of $\chi^2/dof$, information criteria AIC and BIC and the confidence level $CL\%$ for select $\Omega_{rad}$ at fixed $w_m=0.000$.}
\begin{tabular}{|c||c|c|c|c|c|c|c|c|}
\toprule 
Criterion & $\chi^2/dof$ & $\chi^2/dof$ & AIC & AIC & BIC & BIC & CL$\%$    & CL$\%$  \\
\colrule
$\Omega_{rad}$ & Model E1 & Model E2 & Model E1 & Model E2 & Model E1 & Model E2 & Model E1 & Model E2  \\
\colrule
0.0000 & 0.5253433 & 0.0000261 & 22.8003 & 9.6673 & 26.6028 & 13.4698 & 76.8994 & 99.9987 \\
0.00126 & 0.5235623 & 0.0000181 & 22.7557 & 9.6671 & 26.5582 & 13.4696 & 76.9679 & 99.9991 \\
0.0050 & 0.5236620 & 0.0000060 & 22.7582 & 9.6668 & 26.5607 & 13.4693 & 76.9641 & 99.9997 \\
0.0055 & 0.5236753 & 0.0000057(22) & 22.7585 & 9.6668 & 26.5611 & 13.4693 & 76.9636 & 99.9997 \\
0.0060 & 0.5236886 & 0.0000057(42) & 22.7589 & 9.6668 & 26.5614 & 13.4693 & 76.9631 & 99.9997 \\
0.0065 & 0.5237022 & 0.0000061 & 22.7592 & 9.6668 & 26.5617 & 13.4693 & 76.9626 & 99.9997 \\
0.00825 & 0.5237495 & 0.0000097 & 22.7604 & 9.6669 & 26.5629 & 13.4694 &
76.9607 & 99.9995 \\
\botrule
\end{tabular}
\label{Table I.}
\end{table}

\begin{table} 
\caption{Models L1 and L2: Values of $\chi^2/dof$, information criteria AIC and BIC and the confidence level $CL\%$ for select $\Omega_{rad}$ at fixed $w_m=0.000$.}
\begin{tabular}{|c||c|c|c|c|c|c|c|c|}
\toprule 
Criterion & $\chi^2/dof$ & $\chi^2/dof$ & AIC & AIC & BIC & BIC & CL$\%$    & CL$\%$  \\
\colrule
$\Omega_{rad}$ & Model L1 & Model L2 & Model L1 & Model L2 & Model L1 & Model L2 & Model L1 & Model L2  \\
\colrule
0.0000 & 0.6849366 & 1.367324 & 22.9548 & 41.3793 & 25.2279 & 43.6524 & 
71.0016 & 50.4765 \\
0.00126 & 0.6929358 & 1.412905 & 23.1708 & 42.6100 & 25.4439 & 44.8830 & 70.7182 & 49.3391 \\
0.0050 & 0.7176244 & 1.552992 & 23.8374 & 46.3923 & 26.1105 & 48.6654 & 
69.8506 & 46.0015 \\
0.0055 & 0.7210327 & 1.572263 & 23.9294 & 46.9127 & 26.2025 & 49.1857 &
69.7316 & 45.5604 \\
0.0060 & 0.7244654 & 1.591663 & 24.0221 & 47.4364 & 26.2952 & 49.7095 & 
69.6120 & 45.1206 \\
0.00825 & 0.7402252 & 1.680540 & 24.4476 & 49.8361 & 26.7207 & 52.1092 & 69.0657 & 43.1594 \\
\botrule
\end{tabular}
\label{Table II.}
\end{table}

\begin{table} 
\caption{Values of $\chi^2/dof$, information criteria AIC and BIC and the confidence level $CL\%$ for Models E1 and E2 for select $w_m$ at fixed $\Omega_{rad}=0.0055$ compared to Model A and $\Lambda$CDM Model.}
\begin{tabular}{|c||c|c|c|c|c|c|c|c|}
\toprule 
Criterion & $\Lambda$CDM Model & Model A & Model E1 & Model E1 & Model E1 & Model E2 & Model E2 & Model E2  \\
\colrule
$w_m$ & 0.00 & NA & -$\frac{1}{3}$ & 0.00 & +$\frac{1}{3}$ & -$\frac{1}{3}$ & 0.00 & +$\frac{1}{3}$  \\
\colrule
$\chi^2/dof$ & 0.7680 & 0.5205 & 0.5236858 & 0.5236758 & 0.5250609 & 0.0000057(38) & 0.0000057(22) & 0.0014605 \\
AIC & 26.9269 & 20.4921 & 22.7588 & 22.7586 & 22.7932 & 9.6668 & 9.6668 & 9.7032 \\
BIC & 30.0688 & 23.6339 & 26.5613 & 26.5611 & 26.5957 & 13.4693 & 13.4693 & 13.5057 \\
CL$\%$ & 68.1145 & 77.0873 & 76.9632 & 76.9636 & 76.9103 & 99.9997 & 99.9997 & 99.9270 \\
\botrule
\end{tabular}
\label{Table III.}
\end{table}

\begin{table} 
\caption{Present values of energy densities and the parameters $A_T$ and $B_T$ for the best fits of the Models E1 and E2 compared with $\Lambda$CDM Model and Model L1. For comparison with the Cyclic Model A see Table VIII. For $w_m=0$ $\Omega_{M,0}$ has the meaning of the pure Dark Matter. For $w_m=w_r=-\frac{1}{3}$ $\Omega_{M,0}$ combines Dark Matter and atomic matter and $\Omega_{m,0}$ has the meaning of the "residual" matter density $\Omega_{r,0}$.}
\begin{tabular}{|c||c|c|c|c|c|c|c|c|c|}
\toprule 
Model & $w_m $ & $\Omega_{0,0}$ & $\Omega_{M,0}$ & $\Omega_{m,0}$ & $\Omega_{rad,0}$ & $A_T$ & $B_T$ \\
\colrule
$\Lambda$CDM & 0 & $0.692\pm0.012$ & 0.258 & 0.0484 & 0.0016 & 0.000 & 0.000 \\
L1 & 0 & 0.717878$\pm$0.031760 & 0.233312$\pm$0.016006 & 0.047571 & 0.00126 & 0.000 & 0.000 \\
E1 & 0 & 0.614470$\pm$0.159190 & 0.335899$\pm$0.148401 & 0.048371 & 0.00126 & 3.54561$\pm$5.00992 & 24.0825$\pm$68.1344 \\
E1 & -$\frac{1}{3}$ & 0.598408$\pm$0.159195 & 0.351962$\pm$0.148406 &  0.048371 & 0.00126 & 3.42012$\pm$5.01026 & 24.9139$\pm$68.1387 \\
E2 & 0 & 0.530592$\pm$0.063365 & 0.415513$\pm$0.049656 & 0.048396 &   0.0055 & 0.564774$\pm$1.68095 & -18.5660$\pm$21.9465 \\
E2 & -$\frac{1}{3}$ & 0.514549$\pm$0.063360 & 0.431557$\pm$0.049652 & 0.048395 & 0.0055 & 0.439135$\pm$1.68083 & -17.7316$\pm$21.9448 \\
\botrule
\end{tabular}
\label{Table IV.}
\end{table}

\begin{figure} [htp]
\includegraphics[width=12cm,height=10.5cm]{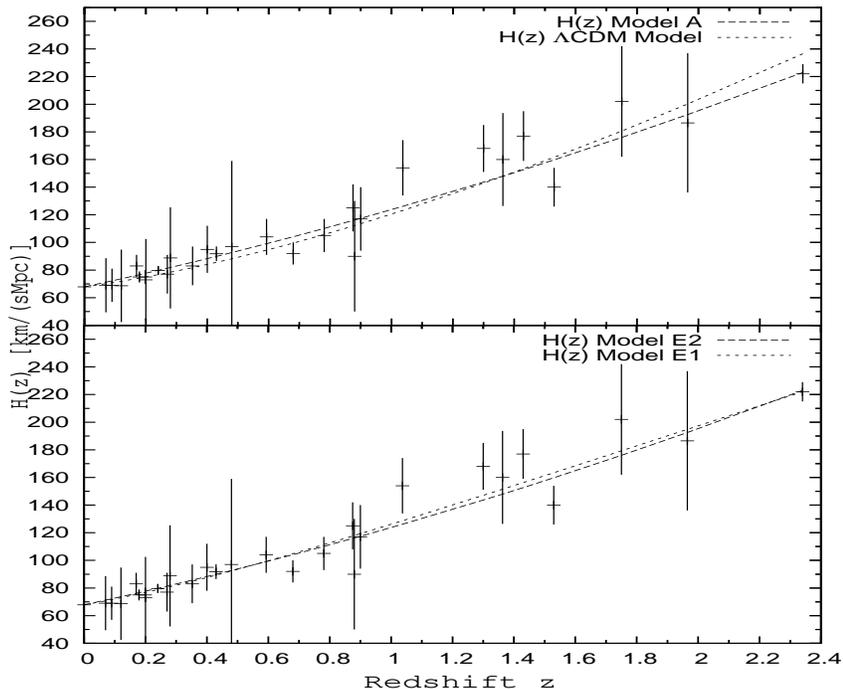}
\caption{Fitted Hubble functions of the Model A and $\Lambda$CDM Model (top) and Models E2 and E1 (bottom) with $w_m=0$ and $\Omega_{rad}=0.0055$ and $0.00126$, respectively, compared to the Hubble data.}
\label{Figure 1}
\end{figure}

\begin{figure} [htp]
\includegraphics[width=12cm,height=10.5cm]{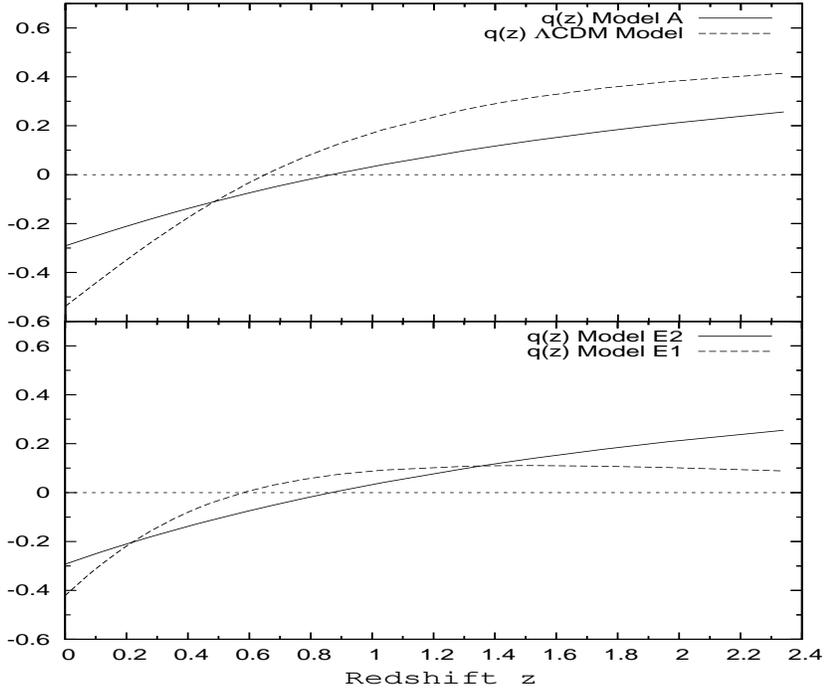}
\caption{Predictions for the deceleration parameter $q(z)$ in the Model A and $\Lambda$CDM Model (top) and in Models E2 and E1 (bottom) with $w_m=0$ and $\Omega_{rad}=0.0055$ and $0.00126$, respectively.}
\label{Figure 2}
\end{figure}
\begin{table} 
\caption{Predictions for the present value of the deceleration parameter $q_0$ and approximate values of the redshift $z_{da}$ at the deceleration-acceleration transition points $q_{da}=0$ in the $\Lambda$CDM Model, Model A and the Models E1 and E2 with $w_m=0$ and with $\Omega_{rad}=0.0055$ and $0.00126$, respectively.}
\begin{tabular}{|c||c|c|c|c|}
\toprule
Model & $\Lambda$CDM & Model A & Model E1 & Model E2 \\
\colrule
$q_0$ & -0.5380 & -0.2912 & -0.4211 & -0.2931 \\
$z_{da}$ & 0.6512 & 0.8667 & 0.5816 & 0.8676 \\
\botrule
\end{tabular}
\label{Table V.}
\end{table}
To select the best model we compared the four models in all runs using the values of $\chi^2/dof=\chi^2_{min}/(N-k)$ where $N$ is the number of data points and $k$ the number of fitted parameters, and the confidence level CL$\%=\exp(-\frac{1}{2}\chi^2/dof)100.0$. In addition we calculated the values of the Akaike and Baysian information criteria, AIC and BIC. These are defined as~\cite{akaike74,schwartz78,burnham02}
\begin{eqnarray}
\text{AIC} & = & \chi^2_{min} +\frac{2kN}{N-k-1}\\
\nonumber
\text{BIC} & = & \chi^2_{min} + k \ln N
\end{eqnarray}
The larger is the difference with respect to the model that carries smaller value of AIC (BIC), the higher is the evidence against the model with larger value of AIC (BIC). 

The results of the first study are shown in the Table I for the Models E1 and E2, and in the Table II for the Models L1 and L2. The values of $\Omega_{rad,0}$ range from the lower limit 0.00126 to the near average value 0.00825. For the comparison we include the value $\Omega_{rad,0}=0.000$. 
We find that the fits to both $H(z)$ and $AH(z)$ data are sensitive to the values of $\Omega_{rad,0}$. The overall best fit of the Models E1 and E2 by $\chi^2/dof$ and all information criteria is for the lower limit of $\Omega_{rad,0}=0.00126$ for the Model E1 and for $\Omega_{rad,0}=0.0055$ for the Model E2. The overall best fit of both Models L1 and L2  by $\chi^2/dof$ and all informaition criteria is for the unphysical value $\Omega_{rad,0}=0.000$. This suggests that these models are unphysical.

The results of the second study are shown in the Table III for the two special values $w_m=-\frac{1}{3}$ and $w_m=0$, and for $w_m=+\frac{1}{3}$
representing $w_m>0$. Compared to $w_m=0$ the values of $\chi^2/dof$ and all information criteria worsen for $w_m>0$ which are therefore excluded. On the other hand, the values of $\chi^2/dof$ and all information criteria remain nearly constant for a broad range of negative $w_m$ from $w_m=-1$ to the special value $w_m=-\frac{1}{3}$ which is therefore not excluded. 

The Table III presents also the results of fits to the $H(z)$ data for the $\Lambda$CDM Model and the Model A. The unphysical Model L1 with $\Omega_{rad,0}=0.000$ has a better $\chi^2/dof$ and all information criteria than $\Lambda$CDM Model but it is excluded by the comparisons with Models A, E1 and E2. The Model L2 is excluded because of the very large values of $\chi^2/dof$, AIC and BIC.

\begin{figure} [htp]
\includegraphics[width=12cm,height=10.5cm]{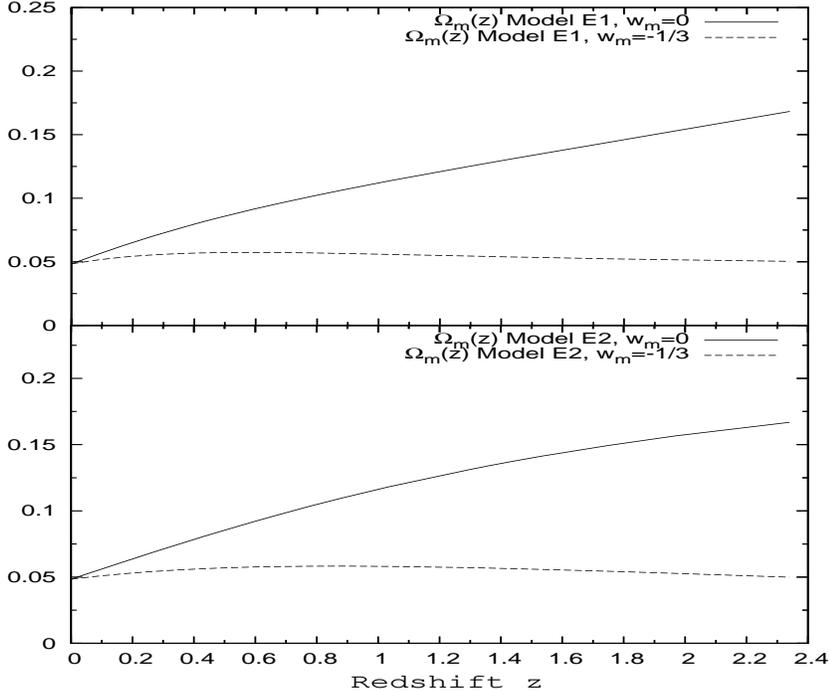}
\caption{Predictions for the normalized fractional energy density of atomic matter $\Omega_m(z)$ with $w_m=0$ and $w_m=-\frac{1}{3}$ in Models E1 (top) and E2 (bottom) with $\Omega_{rad}=0.00126$ and $0.0055$, respectively.}
\label{Figure 3}
\end{figure}

The fit of the Model A with $\chi^2/dof = 0.5205$ is substatially better than the fit of the $\Lambda$CDM Model with $\chi^2/dof=0.7680$. With $\chi^2/dof \sim 0.5205$ all the fits of the Model E1 are very close to the Model A with the closest fit being for $\Omega_{rad,0}=0.00126$. With astonishing small values of $\chi^2/dof$ all fits of the Model E2 to the $AH(z)$ data are nearly identical to the Model A with the closest fit being for $\Omega_{rad,0}=0.0055$ and $\chi^2/dof=0.0000057$. 

Figure 1 shows the best fits of the Model A and the Models E1 and E2 to the Hubble data $H(z)$. All these Models have a better fit at large and smaller $z$ in comparison with the $\Lambda$CDM model. Models E1 and E2 differ slightly at $z\sim 1.0-2.0$. As expected the Models A and E2 are nearly identical. The fitted values of the present energy densities $\Omega_{0,0}$ and $\Omega_{M,0}$ of Dark Energy and Dark Matter and the entropic parameters $A_T$ and $B_T$ are given in the Table IV for both $w_m=0$ and $w_m=-\frac{1}{3}$. To facilitate the reproduction of our results we present these parameters with the original precision of six decimal digits.

The results for $\Omega_{0,0}$ and $\Omega_{M,0}$ for Model E2 are identical to the predictions (3.10) for Model A from relations (3.9) in Section III.B. The values of $\Omega_{0,0}(\Omega_{M,0})$ are smaller (larger) compared to the $\Lambda$CDM Model for both cases $w_m=$ and $w_m=-\frac{1}{3}$. The values of $\Omega_{0,0}(\Omega_{M,0})$ are larger (smaller) for $w_m=0$ compared to $w_m=-\frac{1}{3}$. These distinctions confirm the constraints (1.3) and (1.4) predicted by our model of quantum spacetime and indicate positive curvature density $\Omega_{c,0}$. 

Figure 2 shows the predictions of the models for the deceleration parameter $q(z)$. All models predict late time deceleration-acceleration transitions. The Table V shows the predictions for present value of the acceleration parameter $q_0$ and the values of the redshift $z_{da}$ at which the transitions occur corresponding to $q_{da}=0$. The results for $w_m=-\frac{1}{3}$ are very similar.

Figure 3 presents the predictions of the Models E1 amd E2 for the normalized fractional energy density of $\Omega_m(z)$ for the two values of $w_m=0$ and $w_m=-\frac{1}{3}$. While the results for E1 and E2 are similar, the differences between the atomic matter with $w_m=0$ and the "residual" mass term with $w_m=-\frac{1}{3}$ are dramatic. This is remarkable since the values of the $\chi^2/dof$ are the same in the separate fits with $w_m=0$ and $w_m=-\frac{1}{3}$.

Our Model A of the Cyclic Universe has the best overall fit to the Hubble data. The results presented in the Table III and in the Figures 1 and 2 indicate that the Entropic Models E1 and especially the Model E2 are nearly identical to the Model A. Therefore they represent the Cyclic Universe {\it and} validate the quantum information model of the Dark Energy and Dark Matter as the quantum structure of the spacetime. To complete our description of the spacetime of the Cyclic Universe we shall use the Hubble functions of these models to determine the spatial curvature.

\section{Analysis of Angular diameter distance data and the results}

\subsection{Etherington relation}

Because the definition of the Hubble parameter involves only the scale factor and not the curvature parameter $R_0$ in the RW metric, the curvature term is absent in the Friedmann equations (2.8) and (2.9) and in the relations (3.9). We can still define 
$\Omega_{c}(t)=\rho_c(t)/\bar{\rho}(t)$ so that
\begin{equation}
\Omega_{c,0}=\frac{-kc^2}{R_0^2 H_0^2 a_0^2}
\end{equation}
where $a_0=a(t_0)$. We determine the Hubble function of Entropic and Non-Entropic Models by fitting Hubble data $H(z)$. With these fitted models we can determine $\Omega_{c,0}$ by fitting the luminosity distance data $d_L(z)$ or angular diameter distance data $d_A(z)$ using the equations~\cite{carroll04}
\begin{eqnarray}
d_L(z) & = & (1+z)\frac{1}{\sqrt{|\Omega_{c,0}}|}S_k
\Bigl[\sqrt{|\Omega_{c,0}}|\int \limits_0^z \frac{dz'}{H(z')} \Bigr]\\
d_A(z) & = & \frac{d_L(z)}{(1+z)^2}
\end{eqnarray}
where $S_k(\chi)= \sin(\chi), \chi,\sinh(\chi)$ apply to closed, flat and open geometry with $k=+1,0,-1$, respectively. The luminosity distance relation (6.2) is a direct consequence of the homogeneity and isotropy of the spacetime~\cite{carroll04}. The angular diameter distance relation (6.3) is called Etherington relation, or cosmic distace-duality relation. It follows from the so called reciprocity relations proved by Etherington in 1933~\cite{etherington33} under three fundamental assumptions~\cite{ellis13}: 
(1) spacetime is described by a Riemannian metric theory of gravity
(2) the source and observer are connected by unique null geodesics
(3) the emitted number of photons is conserved.
The validity of the Etherington relation was recently probed by several groups~\cite{ellis13,Lv16,holanda16a,holanda16b,yang17,wang17,li17a,li17b,avgoustidis09}.  Deviations from (6.3) are described by a parameter
\begin{equation}
\eta(z) = \frac{d_L(z)}{(1+z)^2d_A(z)}
\end{equation}
Etherington relation requires $\eta(z)\equiv 1$. The condition (3) can be violated by non-conservation of the photon number by any source of attenuation, e.g. by interstellar dust, gas or plasme or by interactions with exotic particles. The condition (1) could be violated by various variants of modified gravity or by a violation of the Equivalence Principle~\cite{ellis13,holanda16a,avgoustidis12}. Within the General relativity the Etherington relation could be also violated by the anisotropy of the universe~\cite{li17a}.

With Taylor expansions of $S_k(\chi)$ and the integral in $\chi$ 
\begin{eqnarray}
d_L(z)=(1+z)\frac{c}{H_0}\sum \limits_{n=0}^\infty \Omega_{c,0}^n 
\frac{1}{(2n+1)!}F^{2n+1}\\
F(z)=\int \limits_0^z \frac{dz'}{H(z')} = \sum \limits_{m=0}^\infty \frac{1}{m!}\frac{d^mF(0)}{dz^m} z^m
\end{eqnarray}
we get to the fifth order
\begin{eqnarray}
d_L(z) & = & (1+z)\frac{c}{H_0}\Bigl[H_0F+\frac{1}{6}\Omega_{c,0}(H_0F)^3 + \frac{1}{120}(H_0F)^5 \Bigr]\\
H_0F & = & F_1z+\frac{1}{2}F_2z^2+\frac{1}{6}F_3z^3+\frac{1}{24}F_4z^4+\frac{1}{120}F_5Z^5\\
\nonumber
(H_0F)^3 & = & F_1^3z^3 + \frac{3}{2}F_1^2F_2z^4+\frac{3}{4}F_1F_2^2z^5 +\frac{1}{2}F_1^2F_3z^5\\
\nonumber
(H_0F)^5 & = & F_1^5 z^5
\end{eqnarray}
The derivatives $F_m=H_0\frac{d^mF(0)}{dz^m}$ are given by the derivatives of $H(z)$ at $z=0$. With $F_1 = 1$ 
\begin{eqnarray}
F_2 & = &-\Biggl(\frac{1}{H_0}\frac{dH}{dz}\Biggr)\\
\nonumber
F_3 & = &2\Biggl(\frac{1}{H_0}\frac{dH}{dz}\Biggr)^2-\Biggl(\frac{1}{H_0}\frac{d^2H}{dz^2}\Biggr)\\
\nonumber
F_4 & = & -6\Biggl(\frac{1}{H_0}\frac{dH}{dz}\Biggr)^3+6\Biggl(\frac{1}{H_0}\frac{dH}{dz}\Biggr)\Biggl(\frac{1}{H_0}\frac{dH^2}{dz^2}\Biggr) -\Biggl(\frac{1}{H_0}\frac{d^3H}{dz^3}\Biggr)\\
\nonumber
\end{eqnarray}
\[
F_5 = 24\Biggl(\frac{1}{H_0}\frac{dH}{dz}\Biggr)^4-36\Biggl(\frac{1}{H_0}\frac{dH}{dz}\Biggr)^2\Biggl(\frac{1}{H_0}\frac{d^2H}{dz^2}\Biggr)+8\Biggl(\frac{1}{H_0}\frac{dH}{dz}\Biggr)
\Biggl(\frac{1}{H_0}\frac{d^3H}{dz^3}\Biggr)
+6\Biggl(\frac{1}{H_0}\frac{d^2H}{dz^2}\Biggr)^2
-\Biggl(\frac{1}{H_0}\frac{d^4H}{dz^4}\Biggr)^4
\]
Numerical values of the parameters $F_k$, $k=1,5$ for the fitted models are presented in the Table XII in the Appendix B.

\subsection{Data analysis and results}

It is evident from the relation (6.2) that the determination of $\Omega_{c,0}$ from $d_A(z)$ data depends crucially on the assumed Hubble function. Since the Hubble functions of the Models A, E1 and E2 differ considerably from the Hubble function of the $\Lambda$CDM Model, we do not expect a flat Universe so typical for the 
$\Lambda$CDM Model. However we do expect the results for the Models E1 and E2 to be similar to the Model A.

Several parametrizations of $\eta(z)$ were introduced in the literature: $\eta(z)=(1+z)^\epsilon$~\cite{wang17,avgoustidis09}, $\eta(z)=1+\eta_0'z$~\cite{holanda16b,yang17} and others~\cite{yang17,wang17,li17a}. Parameter $\epsilon$ is referred to as opacity. We shall use the form
\begin{equation}
\eta(z)=\frac{1}{1+\eta_0z}
\end{equation}
For small $\epsilon$ and $\eta_0'$ we recover the forms of Ref.~\cite{wang17,avgoustidis09} and Ref.~\cite{holanda16b,yang17} with $\eta_0=-\epsilon$ and $\eta_0=-\eta_0'$, respectively. We shall work with a rescaled angular diameter distance $d_A(z,\eta_0)$
\begin{equation}
d_A(z,\eta_0)=\frac{d_{A,obs}(z)}{1+\eta_0z}=\frac{d_L(z)}{(1+z)^2}
\end{equation}
where $d_{A,obs}(z)$ are the measured data. The measured data from Ref.~\cite{filippis05}-Ref.~\cite{anderson14} and the rescaled data are presented in the Table XI in the Appendix A. In addition we work also with data $Ad_A(z,\eta_0)$ predicted by the best fit of Model A to data $d_A(z,\eta_0)$. The data $d_A(z,\eta_0)$ were fitted by $\Lambda$CDM Model, Model A, Models E1 and E2 with $w_m=0$ and $-\frac{1}{3}$, and by models L1 and L2 with $w_m=0$. The data $Ad_A(z,\eta_0)$ were fitted by the Models E2 and L2. To calculate the curvature parameter $R_0$ from (6.1) we used the present value of the scale parameter from the Model A 
\begin{table} 
\caption{Predictions for the present value of the curvature density $\Omega_{c,0}$ and the curvature parameter $R_0$ from the fits of the $\Lambda$CDM Model, Model A, Models E1 and E2 with $w_m=0$ and $w_m=-\frac{1}{3}$ and Models L1 and L2 with $w_m=0$ to the measured angular diameter distance data $d_A(z)$ and to the rescaled measured data $d_A(z,\eta_0)$ assuming a correction factor $\eta_0=0.0150$. Since the AIC and BIC information criteria follow the $\chi^2/dof$ values, only $\chi^2/dof$ and CL$\%$ are shown to compare the models.}
\begin{tabular}{|c||c|c|c|c|c|c|c|c|c|}
\toprule
Model & $\eta_0$ & $\Lambda$CDM & Model A & Model E1 & Model E1 & Model E2 & Model E2 & Model L1 & Model L2 \\
\colrule
$w_m$ &    & 0 & NA & 0 & -$\frac{1}{3}$ & 0 & -$\frac{1}{3}$ & 0 & 0 \\
$\Omega_{rad,0}$ &  & 0 & NA & 0.00126 & 0.00126 & 0.0055 & 0.0055 & 0.0055 & 0.0055 \\
\colrule
$\Omega_{c,0}$ & 0 & 0.227914 & 1.032532 & 1.052533 & 1.064925 & 1.085066 & 1.096195 & -0.090571 & 0.023108 \\
$\sigma_{\Omega_{c,0}}$  & 0 & 0.265023 & 0.239261 & 0.282056 & 0.281916 & 0.240260  & 0.240187 & 0.259042 & 0.261449 \\
$R_0$ $(Glyr)$ & 0 & 21.0519 & 9.89065 & 9.79608 & 9.73891 & 9.64822 & 9.59912 & 33.3950 & 66.1147 \\
$\chi^2/dof$ & 0 & 1.038363 & 1.030324  & 1.028960  & 1.028936 & 1.028782 & 1.028542 & 1.043562 & 1.041653 \\
CL$\%$ & 0 & 59.5007 & 59.7404 & 59.7811 & 59.7819 & 59.7865 & 59.7936 & 59.3463 & 59.4029 \\
\colrule
$\Omega_{c,0}$ & 0.0150 & 0.004000 & 0.831943 & 0.814721 & 0.827228 & 0.883609 & 0.894803 & -0.307513 & -0.196778 \\
$\sigma_{\Omega_{c,0}}$ & 0.0150 & 0.265989 & 0.239578 & 0.283538 & 0.283394 & 0.240600 & 0.240524 & 0.259817 & 0.262301 \\
$R_0$ $(Glyr)$ & 0.0150 & 158.908 & 11.0187 & 11.1344 & 11.0499 & 10.6917 & 10.6246 & 18.1236 & 22.6562 \\
$\chi^2/dof$ & 0.0150 & 1.041249 & 1.029450  & 1.030800 & 1.030804 & 1.028181 & 1.027984 & 1.046931 & 1.044848 \\
CL$\%$ & 0.0150 & 59.4150 & 59.7665 & 59.7262 & 59.7261 & 59.8044 & 59.8103 & 59.2464 & 59.3081 \\
\botrule
\end{tabular}
\label{Table VI.}
\end{table}
\begin{table} 
\caption{Predictions for the present value of the curvature density $\Omega_{c,0}$ and the curvature parameter $R_0$ from the fits of the Model E2 with $w_m=0$ and $w_m=-\frac{1}{3}$ to the angular diameter distance data $Ad_A(z)$ and $Ad_A(z,\eta_0)$ assuming a correction factor $\eta_0=0.0150$. The data $Ad_A(z)$ and $Ad_A(z,\eta_0)$ are the angular diameter distance predicted by the fits of the Model A to the measured angular diameter distance data $d_A(z)$ and to the rescaled measured data $d_A(z,\eta_0)$, respectively. These fits indicate how close is the Model E2 to the Model A.}
\begin{tabular}{|c||c|c|c|c|}
\toprule
Model & Model E2 & Model E2 & Model E2 & Model E2 \\
\colrule
$\eta_0$ & 0 & 0 & 0.0150 & 0.0150 \\
\colrule
$w_m$ & 0 & -$\frac{1}{3}$ & 0 & -$\frac{1}{3}$ \\
\colrule
$\Omega_{c,0}$ & 1.030790 & 1.033490 & 0.829892 & 0.832980 \\
$\sigma_{\Omega_{c,0}}$ & 0.042396 & 0.042397 & 0.042453 & 0.042454 \\
$R_0$ $(Glyr)$ & 9.37181 & 9.36099 & 10.3089 & 10.2944 \\
$\chi^2/dof$ & 0.051478 & 0.064393 & 0.051156 & 0.063754 \\
CL$\%$ & 97.4590 & 96.8316 & 97.4747 & 96.8626 \\
\botrule
\end{tabular}
\label{Table VII.}
\end{table}
$a_0=a(t_0)=1.434732$~\cite{svec17a} for all fitted models. The results for $\Omega_{c,0}$ and $R_0$ are shown in the Tables VI and VII. 

Our initial fits assumed no rescaling with $\eta_0=0$. The Table VI shows that in this case it is the Models L1 and L2 which are compatible with a flat spacetime with $\Omega_{c,0}=0$ and not the 
\begin{figure} [htp]
\includegraphics[width=12cm,height=10.5cm]{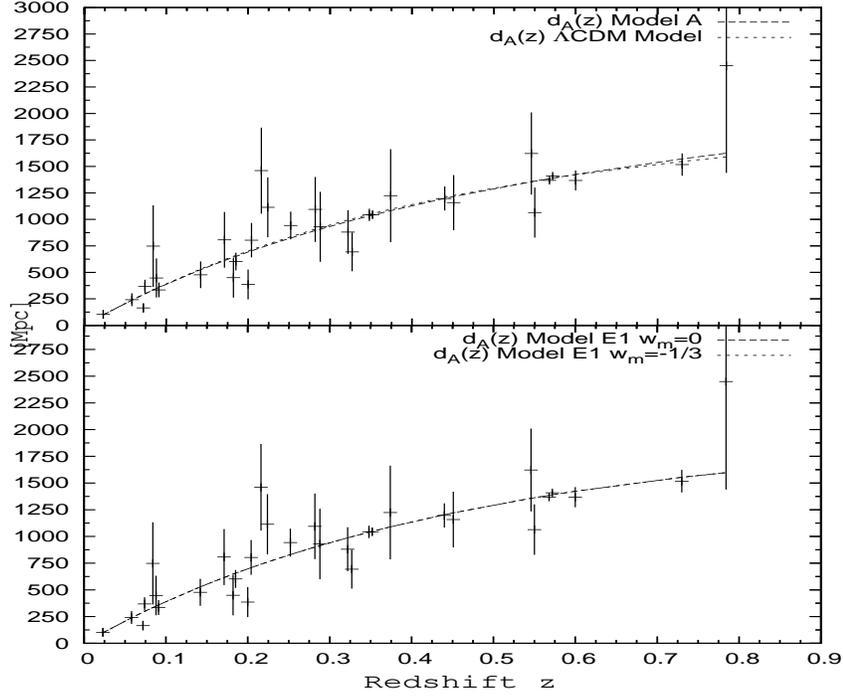}
\caption{Fits of Models A, $\Lambda$CDM (top) and E1 (bottom) to data $d_A(z)$ assuming $\eta_0=0.0150$.}
\label{Figure 4}
\end{figure}
\begin{figure} [hp]
\includegraphics[width=12cm,height=10.5cm]{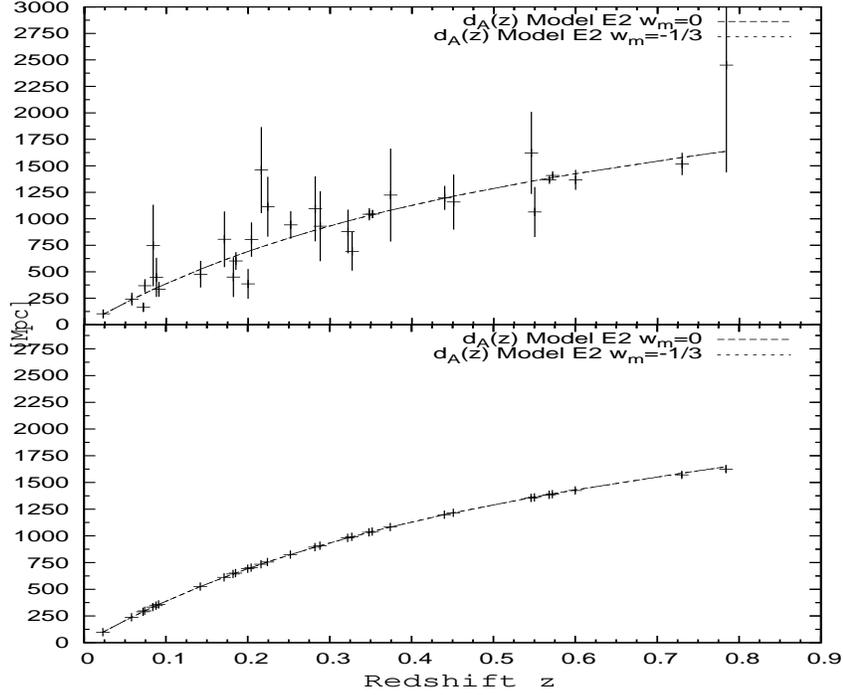}
\caption{Fits of the Model E2 to data $d_A(z)$ (top) and $Ad_A(z)$ (bottom) assuming $\eta_0=0.0150$.}
\label{Figure 5}
\end{figure}
$\Lambda$CDM Model which has a large curvature parameter $\Omega_{c,0}=0.227914$. Since $\Omega_{c,0}(\Lambda$CDM)=0 has been reliably determined by the luminosity distance data and CMB data~\cite{planck15} we need to rescale the $d_{A,obs}(z)$ data. We found that with $\eta_0=0.0150$ we obtain $\Omega_{c,0}(\Lambda$CDM)=0.004000$\pm$0.265989 which is compatible with zero. The large errors are in agreement with recent analysis of the $H(z)$ and BAO data~\cite{yu17}. All models were then refitted to the rescaled data with this $\eta_0$. Our value for $\eta_0$ is in excellent agreement with the very recent model independent test of the Etherington relation (6.11) which yields $\eta_0=0.0147^{+0.056}_{-0.066}$~\cite{ruan18}.

The results in the Table VI show that both fits share certain systematic patterns. For both $\eta_0$ all models have a very similar $\chi^2/dof$. Still, Model E2 with $w_r=-\frac{1}{3}$ has the best CL$\%$,followed closely by Model E2 with $w_r=0$ and Model A. The  $\Lambda$CDM Model has appreciably lower CL$\%$ than Model A, but still higher than the Models L1 and L2. The results with $w_m=0$ and $-\frac{1}{3}$ are esentially identical. This is confirmed in the Figures 4 and 5 which show the predicted rescaled angular diameter distance for all fits with $\eta_0=0.0150$. The results in the Table VII show that Model E2 is nearly identical to the Model A. Importantly, this study shows that the entropic models share similar large value of curvature density $\Omega_{c,0}$ in agreement with recent analysis~\cite{wang17}. 

The angular diameter distance data confirm the positivity of the curvature density $\Omega_{c,0}$ and the validity of the constraints (1.3) and (1.4) on the cosmological parameters. Together with the analysis of the Hubble data these results indicate that the Entropic Models E1 and E2 represent the Cyclic Universe and validate the quantum information model of Dark Energy and Dark Matter as the quantum structure of the spacetime. However the analysis adds new information: the spacetime is an open anti-de Sitter spacetime.

\section{Past and future evolution of the Cyclic Universe.}

The central result of the Friedmann equations supplemented by the Laws of Thermodynamics is the universal entropic form (1.1) of the Hubble function which predicts the existence of dynamical Dark Energy and dynamical Dark Matter with equations of state $w_0=-1$ and $w_M=0$, respectively, at all $z$ and all times $t$. The Hubble function $H(A)$ of the Model A is constructed to describe a Cyclic Universe and it should be identified with its equivalent entropic form of Model E so that $H(E)\equiv H(A)$. 

We have seen that the fits of Model E to the Model A (called Model E2) at $z=0-2.34$ render Model E2 essentially identical to the Model A. These fits assume an approximation to the entropic terms $\Sigma_0(z)$ and $\Sigma_M(z)$ based on our model of Dark Energy and Dark Matter as the quantum spacetime. For $z\gtrsim 2.5$ the fits begin to deviate slightly from the Model A, indicating the limits of the approximation. However, we can identify the Model E2 with Model A at all $z$ and all $t$ and use Friedmann equations to calculate the entropic terms beyond this approximation. This is possible since we know the Hubble function of the Model A as a function of the redshift $z$  as well as a function of the cosmic time $t$, and since we know all non-entropic terms of the Hubble function of the Model E2 from the fits at $z=0-2.34$.

The Hubble function $H(z)$ is given by (3.1). The Hubble function $H(t)$ is given by (3.4). With the Hubble functions known we can calculate the deceleration parameters $q(z)$ and $q(t)$. From the Friedmann equations (3.8) we then determine $\Omega_0(z)$ and $\Omega_M(z)$, and similarly $\Omega_0(t)$ and $\Omega_M(t)$. Given by (3.9), we can split these fractional energy densities into non-entropic and entropic terms
\begin{eqnarray}
\Omega_0 & = & \Omega_{00} + \Omega_{0S}= \Bigl[\Omega_{0,0}+\Sigma_0\Bigr]\frac{H_0^2}{H^2}\\
\nonumber
\Omega_M & = & \Omega_{M0} + \Omega_{MS}= \Bigl(\frac{a(t_0)}{a(t)}\Bigr)^3\Bigl[\Omega_{M,0}+\Sigma_M\Bigr]\frac{H_0^2}{H^2}
\end{eqnarray}
where $\frac{a(t_0)}{a(t)}=1+z$ for the $z$-dependence and $a(t)=\frac{F^n(t)}{1-CF^n(t)}$ for the cosmic time dependence~\cite{svec17a}. The non-entropic terms $\Omega_{00}$ and $\Omega_{M0}$ are known from the fit E2 and the entropic terms $\Omega_{0S}$ ($\Sigma_0$) and $\Omega_{MS}$ ($\Sigma_M$) can be determined from (7.1). The results for the past evolution ($z$-dependence) and for the future evolution (t-dependence) of $\Omega_0$ and $\Omega_M$ are presented, discussed and compared 
\begin{figure} [htp]
\includegraphics[width=12cm,height=10.5cm]{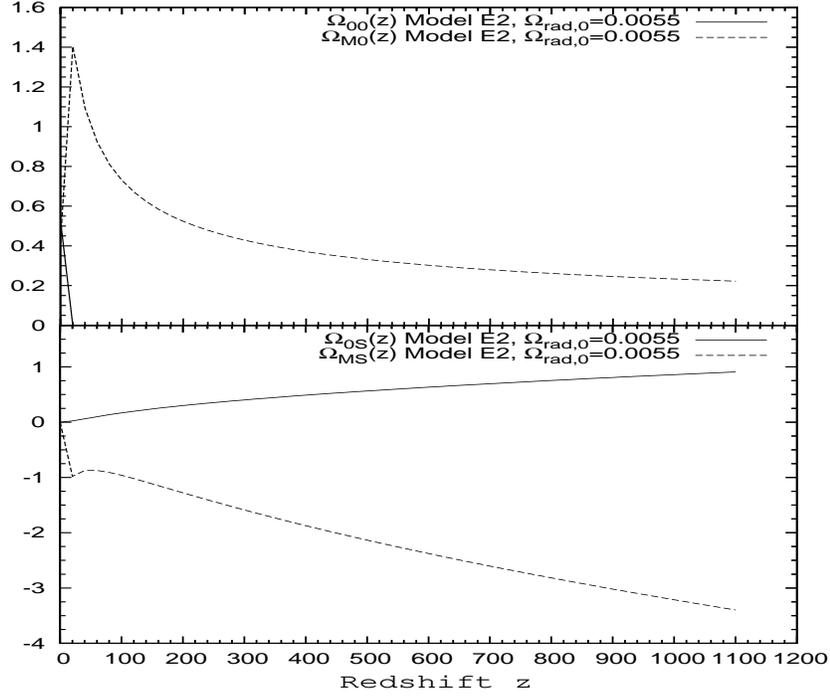}
\caption{Non-Entropic (top) and Entropic (bottom) terms of Dark Energy and Dark Matter at high $z$.}
\label{Figure 6}
\end{figure}

\begin{figure} [hp]
\includegraphics[width=12cm,height=10.5cm]{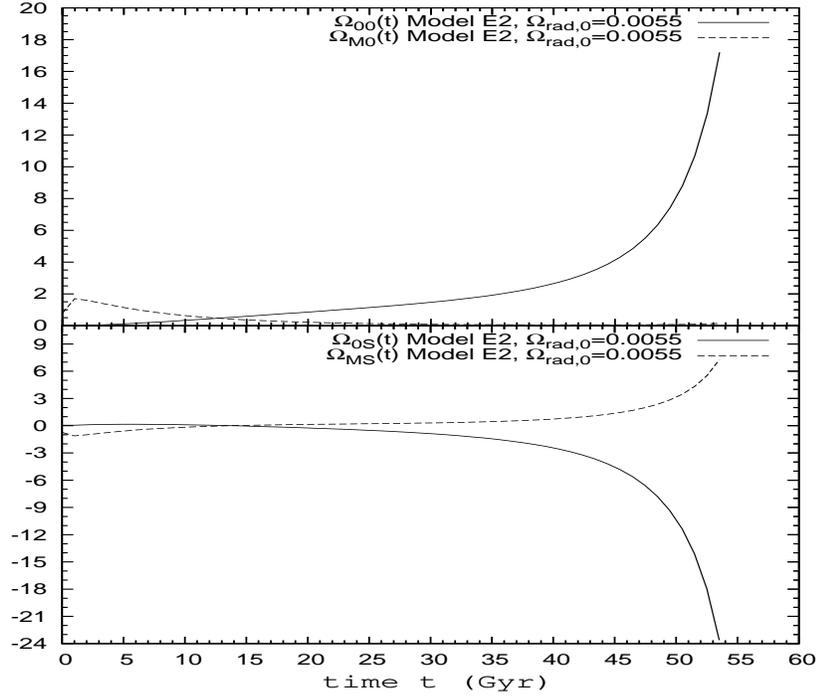}
\caption{Time evolution of Non-Entropic and Entropic terms of Dark Energy and Dark Matter.}
\label{Figure 7}
\end{figure}
to the $\Lambda$CDM Model in Ref.~\cite{svec17a}. The results for the past and future evolution of $\Omega_{0S}$ and $\Omega_{MS}$ are shown in Figures 6 and 7. It is noteworthy that at $z=0/t=t_0$ the solutions $\Omega_0$ and $\Omega_M$ 
determine the exact present values $\Omega_{0,0}$ and $\Omega_{M,0}$ in the Model A
\begin{eqnarray}
\Omega_0(t_0) & = & \Omega_{00}(t_0)=\Omega_{0,0}\\
\nonumber
\Omega_M(t_0) & = & \Omega_{M0}(t_0)=\Omega_{M,0}
\end{eqnarray}
The exact values of $\Omega_{0,0}$ and $\Omega_{M,0}$ in the Model A are shown and compared with the fits of the Model E2 in the Table VIII for $w_m=0$ and $w_m=-\frac{1}{3}$. The results show that the values from the fits of the Model E2 are essentially identical to the values from the Model A for either value of $w_m$ providing evidence that these two Models are identical. The Table VIII shows that there is a small difference between the solutions with different $w_m$. 

We can also compare the evolution of $\Omega_0(z)$ and $\Omega_M(z)$ in the exact Model A and in the approximate Model E2 to see how far we can assume the "small" $z$ approximation of $\Sigma_0(z)$ and $\Sigma_M(z)$ in the Model E2. Results at selected values of $z$ are shown in the Table IX for $w_m=0$. The results show a full agreement up to $z\approx 3.0$ and only a small difference at $z=5.0$. At $z=20.0$ $\Omega_0=0.02272$ for Model A and 0.01843 for Model E2 while $\Omega_M=0.42357$ for Model A and 0.40640 for Model E2. This diference increases rapidly with increasing $z$. Above $z\sim 20$ $\Omega_0(z)$ is increasing in Model A while it is rapidly decreasing in Model E2. Above $z\sim 20$ $\Omega_M(z)$ is decreasing becoming negative for $z \geq 70$ in Model A while it is positive and decreasing in Model E2. We conclude that the approximation used in the Model E2 works reasonably well for $z<20$ but breaks down completely for $z>20$. 

\begin{table}
\caption{Present values of $\Omega_{0,0}$ and $\Omega_{M,0}$ predicted by the Model A compared to the fitted values in Model E2 for $w_m=0$ and $w_m=-\frac{1}{3}$. For $w_m=0$ $\Omega_{M,0}$ has the meaning of the pure Dark Matter. For $w_m=w_r=-\frac{1}{3}$ $\Omega_{M,0}$ combines Dark Matter and atomic matter and $\Omega_{m,0}$ has the meaning of the "residual" matter density $\Omega_{r,0}$.}
\begin{tabular}{|c|c|c|c|c|c|c|c|c|}
\toprule 
$w_m$ & 0 & 0 & 0 & 0 & -$\frac{1}{3}$ & -$\frac{1}{3}$ & -$\frac{1}{3}$ & -$\frac{1}{3}$\\
\colrule
Model & Model A & Model E2 & Model A & Model E2 & Model A & Model E2 & Model A & Model E2 \\
\colrule
      & $\Omega_{0,0}$ & $\Omega_{0,0}$ &  $\Omega_{M,0}$ & $\Omega_{M,0}$ &   $\Omega_{0,0}$ & $\Omega_{0,0}$ &  $\Omega_{M,0}$ & $\Omega_{M,0}$ \\
\colrule
      & 0.529291 & 0.530592 & 0.416809 & 0.415513 & 0.513157 &  0.514549 & 0.432943 & 0.431557 \\
\botrule
\end{tabular}
\label{Table VIII.}
\end{table}

We have arrived at the entropic form (1.1) for the Model A: the non-entropic terms $\Omega_{0,0}$ and $\Omega_{M,0}$ are given by (7.2) and the entropic terms $\Sigma_0$ and $\Sigma_M$ by (7.1). The atomic matter/"residual" mass terms and the radiation terms are the same as in Model E2. We identify the exact (no approximation) Entropic Model E2 with the Model A at all $z$ and at all $t$ using the relations (7.1). The exact entropic terms $\Sigma_0$ and $\Sigma_M$ in (7.1) satisfy the relation
\begin{equation}
\frac{d\Sigma_0}{da}+\Bigl(\frac{a_0}{a}\Bigr)^3 \frac{d\Sigma_M}{da}=0
\end{equation}
as the result of the conservation of the entropy of Dark Energy and Dark Matter $S_0(t)+S_M(t)=S'=const$.

\begin{table}
\caption{Comparison of the evolution of the exact $\Omega_0(z)$ and $\Omega_M(z)$ predicted by Model A with the approximate solutions for $\Omega_0(z)$ and $\Omega_M(z)$ predicted by Model E2 at selected values of $z$ for $w_m=0$. The results with $w_m=-\frac{1}{3}$ are similar.}
\begin{tabular}{|c|c|c|c|c|c|c|c|c|c|c|}
\toprule
$z$ & 1.0 & 1.0 & 2.0 & 2.0 & 3.0 & 3.0 & 4.0 & 4.0 & 5.0 & 5.0 \\
\colrule
Model & Model A & Model E2 & Model A & Model E2 & Model A & Model E2 & Model A & Model E2 & Model A & Model E2 \\
\colrule
$\Omega_0(z)$ & 0.320266 & 0.320326 & 0.208976 & 0.209214 & 0.144814 & 0.148021 & 0.104725 & 0.111754 & 0.078231 & 0.088588 \\
$\Omega_M(z)$ & 0.537054 & 0.536974 & 0.579755 & 0.579511 & 0.593577 & 0.590029 & 0.595223 & 0.586803 & 0.590938 & 0.577512 \\
\botrule
\end{tabular}
\label{Table IX.}
\end{table}

\section{What is the Cyclic Universe with anti-de Sitter spacetime?}

\subsection{The quantum origin of the spatial curvature}

The average $\Omega_{c,0}$ of the Model A and all entropic models in the Table VI is $(\Omega_{c,0})_{av}=1.0663\pm0.2567$ for $\eta_0=0$ and $(\Omega_{c,0})_{av}=0.8505\pm0.2575$ for $\eta_0=0.0150$. Within errors  all these models are consistent with these averages. $\Lambda$CDM Model definitely favours flat spacetime with $\Omega_{c,0}=0$. Since the Cyclic Model A and the entropic models all have significantly better fits to the Hubble data and even to the angular diameter distance data than the $\Lambda$CDM Model we must seriously consider the possibility that the  space has a negative space curvature $k=-1$. This is a hyperbolic topology on an infinite 3-dimensional space with a curvature parameter $R_0$ which defines an infinite 3+1 dimensional anti-de Sitter (AdS) spacetime. This raises an important question: is there a connection between the curvature parameter $R_0$ and the quantum structure of the spacetime?

We view the fits of the Hubble functions to the rescaled data with $\eta_0=0.0150$ as the physical solutions. The $\Lambda$CDM Model has Hubble function $H^2(\Lambda$CDM) with no entropic terms and has $\Omega_{c,0}=0$. The entropic Models E1 and E2 have Hubble functions $H^2(E)$ with entropic terms and have a large $\Omega_{c,0}(E)>0$. We can write $H^2(E)$=$H^2(\Lambda$CDM)+$\Delta H^2(E)$ where
\begin{eqnarray}
\Delta H^2(E) & = & \Delta H^2(0)+\Delta H^2(\Sigma)\\
\Delta H^2(0) & = & \Omega_{0,0}(E)-\Omega_{0,0}(\Lambda CDM)+
(1+z)^3\Bigl[\Omega_{Mm,0}(E)-\Omega_{Mm,0}(\Lambda CDM)\Bigr]\\
\Delta H^2(\Sigma) & = & \Sigma_0(z) + (1+z)^3\Sigma_M(z)
\end{eqnarray}
where we assume $\Omega_{rad,0}(E)=\Omega_{rad,0}(\Lambda CDM)$. The entropic terms $\Sigma_0(z)$ and $(1+z)^3\Sigma_M(z)$ are determined by our entropic model of the quantum spacetime by the relations (4.6)-(4.9). While these terms vanish at $z=0$ their derivatives at $z=0$ are all non-zero with significant values. The parameters $\Omega_{0,0}(E)$ and $\Omega_{M,0}(E)$ are also determined by $\Sigma_0(z)$ and $(1+z)^3\Sigma_M(z)$ in the fits to Hubble data and have sigificant derivatives at $z=0$. To see the role of these derivatives we express the derivatives $\frac{d^mH}{dz_m}$ in (6.9) in terms of the derivatives of $H^2$ since the entropic as well as non-entropic models are defined in term of $H^2(z)$. With $H^2\equiv Q$ we have
\begin{eqnarray}
\frac{dH}{dz} & = & +\frac{1}{2}\frac{1}{H}\frac{dQ}{dz}\\
\nonumber
\frac{d^2H}{dz^2} & = & 
-\frac{1}{4}\frac{1}{H^3}\Bigl(\frac{dQ}{dz}\Bigr)^2
+\frac{1}{2}\frac{1}{H}\frac{d^2Q}{dz^2}\\
\nonumber
\frac{d^3H}{dz^3} & = & 
+\frac{3}{8}\frac{1}{H^5}\Bigl(\frac{dQ}{dz}\Bigr)^3
-\frac{3}{4}\frac{1}{H^3}\frac{dQ}{dz}\frac{d^2Q}{dz^2}
+\frac{1}{8}\frac{1}{H}\frac{d^3Q}{dz^3}\\
\nonumber
\frac{d^4H}{dz^4} & = &
-\frac{15}{16}\frac{1}{H^7}\Bigl(\frac{dQ}{dz}\Bigr)^4
+\frac{9}{4}\frac{1}{H^5}\Bigl(\frac{dQ}{dz}\Bigr)^2\frac{d^2Q}{dz^2}
-\frac{3}{4}\frac{1}{H^3}\Bigl(\frac{d^2Q}{dz^2}\Bigr)^2
-\frac{1}{H^3}\frac{dQ}{dz}\frac{d^3Q}{dz^3}
+\frac{1}{2}\frac{1}{H}\frac{d^4Q}{dz^4}
\end{eqnarray}

We see that the derivatives $\frac{d^mH}{dz^m}$ depend on powers of all derivatives of $\Delta H^2$ up to the order $m$. The equations (6.9) then show that the derivatives of $\Delta H^2(E)$ up to the order $k-1$ play a crucial role in the determination of the parameters $F_k$ which determine the fitted value of $\Omega_{c,0}$. Since the entire term $\Delta H^2(E)$ originates in our model of the quantum structure of the spacetime the entire difference between the curvature parameter $\Omega_{c,0}$ of the entropic models - and thus the Model A - and the zero curvature parameter of the $\Lambda$CDM Model has the same quantum origin. In an amazing display of the selfconsitency of the entropic model of the Cyclic Universe and its unity the curvature parameter $R_0$ originates in the quantum structure of the spacetime.

\subsection{From Cyclic Universe to a Cyclic Multiverse}

According to the Standard Model, the expanding spacetime of the Universe started with a point-like singularity some 13.81 Gyr ago. The spacetime may continue to expand forever, may break down or collapse on itself in a Big Crunch. Cyclic models suggest that the Universe starts with a very small finite volume $V_i$ of the space and expands to a maximum volume $V_f$ when it turns to a contraction back to the initial volume.

To reconcile the idea of an expanding finite 3-dimensional space with the evidence for an infinite 3-dimensional space with curvature we identify the expanding volume $V(t)$ of the Cyclic Universe with the proper volume
\begin{equation}
V(t)=a(t)^3 \mathcal{V}
\end{equation}
where $\mathcal{V}$ is a finite comoving volume. It follows that
\begin{equation}
\frac{\dot{V}}{V}=3\frac{\dot{a}}{a}=3H
\end{equation}
so that the volume is a periodic function given by 
\begin{equation}
V(t)=V_{min} \exp(3\int \limits_{t_\alpha}^t dt'H(t'))
\end{equation}
where $\int \limits_{t_\alpha}^{t_\alpha+T} dt' H(t')=0$. Then $V(t_\alpha)=V_{min}=a_{min}^3\mathcal{V}$, $V(t_\omega)=V_{max}=a_{max}^3\mathcal{V}$ and $V(t_\alpha+T)=V_{min}$. The cyclic evolution is thus free of initial and final state singularities.

We imagine the infinite comoving space as a lattice where each cell has the same comoving volume $\mathcal{V}$. Irrespective of its shape, this universal volume is given by the curvature parameter $\mathcal{V}=R_0^3$.
We define a proper extent of the Cyclic Universe by $R(t)=(V(t))^{\frac{1}{3}}=a(t)R_0$. Then
\begin{equation}
1+z=\frac{a(t_0)}{a(t)}=\frac{R(t_0)}{R(t)}
\end{equation}
relates the proper extent to the redshift. The expansion velocity of the Universe is given by $dR/dt=H(t)R(t)$. The proper extent and the expansion velocity are observable in the Model A where we know the scale factor $a(t)$ and the Hubble function $H(t)$~\cite{svec17a} as well as $R_0$. The initial, present and final proper extent and expansion velocity of the observable Universe in the Model A are summarized in the Table X. For a more detailed numerical analysis of the evolution of $a(t)$, $R(t)$, $dR/dt$ as well as pressure $\bar{p}(t)$ and curvature $\Omega_c(t)$ see Ref.~\cite{svec17a}.
\begin{table}
\caption{Proper extent and expansion velocity of the Cyclic Universe in Model A assuming $\eta_0=0.0150$.}
\begin{tabular}{|c|c|c|c|}
\toprule 
$time$ $t$ $(Gyr)$ & $a(t)$ & $R(t)$ & $dR/dt$ \\
\colrule
$t_\alpha=$0 & 2.755961x10$^{-32}$ & 2.876777 $\mu m$  & 0 \\
\colrule
$t_0=$13.80 & 1.434732 & 15.808882 $Glyr$ & 1.105594 $c$ \\
\colrule
$t_\omega=$60.586 & 6.834258 & 75.304639 $Glyr$ & 0 \\
\botrule
\end{tabular}
\label{Table X.}
\end{table}

At every point of the infinite space the Robertson-Walker metric is the same with the same value of the periodic scale factor $a(t)$ with a period $T$, and with the same $R_0$. At any time $t$ the scale factor describes the instantaneous state of the evolution of the entire anti-de Sitter spacetime. The state of the evolution at time $t$ is the same at every point of the space as the entire infinite space moves along the time axis. The proposed lattice structure of the cyclic Anti-de Sitter spacetime forms a kind of Cyclic Lattice Multiverse with each cell being a Cyclic Universe. This picture is somewhat akin to the Many Worlds Hypothesis of the Quantum Mechanics.

For the flat spacetime $R_0=\infty$ so that $\mathcal{V}=\infty$. As the result all entropic terms $\frac{kT}{V}\frac{dS_k}{dt}$ and $\frac{\mu_k}{V}\frac{dN_k}{dt}$ vanish and we naturally recover the Non-entropic Models including the $\Lambda$CDM Model. 

\subsection{Lorentz symmetry and the conservation of energy in Cyclic Universe}

With proper coordinates $x^i_p(t)=a(t)x^i$ the Robertson-Walker metric (1.1) takes the form at the present time 
\begin{eqnarray}
g_{ij} & = & \Bigl(\delta_{ij}-\frac{1}{R_0^2}\frac{x^i_p(t_0)x^j_p(t_0)}{1+\frac{1}{R_0^2}|\vec{x_p}(t_0)|^2} \Bigr)\\
\nonumber
g_{i0} & = & 0, g_{00}=-1
\end{eqnarray}
The curvature introduces deviations from the Minkowski metric $\eta_{\mu \nu}=diag (-1.+1,+1,+1)$ which violate Lorentz symmetry. The degree of the violation $\xi=\frac{1}{R_0^2}|\vec{x_p}(t_0)|^2$ depends on the scale of the proper distance from the origin where the observer is located. 

In the Model A $R_0=11.0187$ Glyr $=3,380.33$ Mpc for $\eta=0.0150$. Assuming a scale of 10kpc for a typical galaxy we find $\xi=0.87518$x10$^{-11}$ which is not detectable. For a galaxy cluster at a scale of 1Mpc $\xi=0.87518$x$10^{-7}$. At a large scale of 100Mpc $\xi=0.87518$x$10^{-3}$ which could be an observable effect. At very large scales of 1000Mpc $\xi=0.87518$x$10^{-1}$ and the violation of Lorentz symmetry will be observable. In the Model A 1000Mpc correspond to $z$=0.33489. These are the scales at which the angular diameter distance was measured and the Etherington relation $\eta=1$ was violated. This suggests that the correction factor $\eta_0=0.0150$ is a signal of the violation of the Lorentz invariance at large scales.

Our physical theories are local theories that assume locally flat Minkowski spacetime in agreement with the Equivalence Principle of the General relativity. They apply to the entire Universe only for the flat spacetime with infinite $R_0$. With finite curvature factor they must be generalized to theories in a curved spacetime. Anti-de Sitter relativity theory was recently formulated by Cot\u{a}escu~\cite{cotaescu17}.

In General relativity, time is part of the coordinate system and in general depends on the position. Then, globally, energy is not generally conserved. The Laws of Thermodynamics like General relativity apply to physical systems of any scale. Since the Cyclic Universe is a closed system in an equilibrium, its total internal energy $U$ must be conserved during every cycle of its evolution. With $dU=-\bar{p}dV=-3H\bar{p}Vdt$ a total differential we then require
\begin{equation}
U(T)-U(0)=\int \limits_0^T dU = 
-\int \limits_0^T 3H\bar{p}Vdt = W_{0 \to \frac{T}{2}}+
W_{\frac{T}{2} \to T}= 0
\end{equation}
The condition (8.10) is satisfied by the Cyclic Universe. With $H(T-t)=H(-t)=-H(t)$, $\bar{p}(T-t)=+\bar{p}(t)$ (eq.(2.10)) and $V(T-t)=V(t)$  (eq.(8.7)) the term $W_{0 \to \frac{T}{2}}$ during the expansion is exactly balanced by $W_{\frac{T}{2} \to T}$ during the contraction. This is a non-local form of energy conservation imposed by the cyclic evolution on the Friedmann continuity equation (2.10), and therefore on the Friedmann equations (2.8) and (2.9) themselves. 

Technically, finite $R_0$ implies extremally small violations of the Equivalence Principle and therefore energy conservation even on the local level. For scales of 1m $\xi=0.91893$x$10^{-52}$. At the proton scale $10^{-15}$m $\xi=0.918993$x$10^{-82}$. Such extremally small degrees of violation have no immediately observable effects. Although small, over long times and across the entire Universe such effects could accumulate and affect the  dynamics of the Cyclic Universe. Of particular concern is the cumulative violation of the conservation of energy in particle scattering processes such as the nucleosynthesis processes in the stars and in the early Universe. The self-consistency of the Cyclic Universe therefore requires that any such cumulative effects balance out over each evolution cycle.

\section{Evolution of the spatial curvature density $\Omega_c(t)$.}

Using the equation (2.3) we define a fractional curvature energy density
\begin{equation}
\Omega_c=\frac{8\pi G}{3c^2 H^2}\rho_c = \frac{-kc^2}{R_0^2a^2H^2}
=\Biggl\{\frac{-kc^2}{R^2_0a_0^2}\frac{(1+z)^2}{H_0^2}\Biggr\}
\frac{H_0^2}{H^2(z)}
=\Bigl\{\Omega_{c,0}(1+z)^2\Bigr\}
\frac{H_0^2}{H^2(z)}
\end{equation}
The curvature parameter $R_0$ is a pure geometrical concept that describes the Robertson-Walker metric. Theoretically, it is completely independent of the scale factor and the Hubble function. Since it enters the Friedmann equations it is cosmologically important. Recently several methods were discussed to determine the present value of the curvature density $\Omega_{c,0}$ by pure geometrical methods~\cite{denissenya18,Qi18} which would determine $R_0$ given the measurements of $H_0$ and $a_0$. We shall  presume that we know the curvature parameter $R_0$ from $\Omega_{c,0}$ determined by the fits of $H(z)$ to the luminosity distance or angular diameter distance data. 

When we know the scale factor $a(t)$ and the Hubble function $H(t)$ we can calculate from (9.1) the curvature density $\rho_c(t)$ and the fractional curvature density $\Omega_c(t)$. Let us suppose a static $\Omega_c(t)=const\neq0$. Then $R_0$ is finite and $a(t)H(t)=const$. Taking a derivative of this equation we find that the deceleration parameter $q(t)=0$ at all times $t$. A vanishing static $\Omega_c(t)=const\equiv0$ requires $R_0=\infty$ and allows for an evolving $q(t)\neq0$. 

Various cosmological observations indicate that the expansion of the Universe is accelerating at the present time. The acceleration of the Universe expansion means that either the curvature density $\Omega_c(t)$ is dynamic and evolves with the cosmic time $t$ (and thus with the redshift $z$), or it is static and vanishing. Anti-de Sitter spacetime and the flat spacetime are both infinite spacetimes but differ in the value of $R_0$: while $R_0$ is finite in the first it is infinite in the second spacetime. 

It is interesting to note that the rate of change of $\Omega_c$ is given by 
\begin{equation}
\frac{d\Omega_c}{dt}=2q(t)H(t)\Omega_c(t)
\end{equation}
At the extrema the deceleration parameter $q(t)=0$ and
\begin{equation}
\frac{d^2\Omega_c}{dt^2}=2H\Omega_c\frac{dq}{dt}
\end{equation}
so that the kind of the extremum of $\Omega_c$ is controlled by the sign of $\frac{dq}{dt}$ at this point. Integrating (9.2) from an initial time $t_i$ to the time $t$ we find the evolution equation
\begin{equation}
\Omega_c(t)=\Omega_c(t_i)\exp \Biggl(\int \limits_{t_i}^t 2q(t')H(t')dt' \Biggr)  
\end{equation}
Let $t_i$ be the transition point from the decelerating to the accelerating expasion. With $q(t)$ increaingly negative the equation (9.4) imples a decreasing $\Omega_c(t)$ for $k=-1$.

\begin{figure} [htp]
\includegraphics[width=12cm,height=10.5cm]{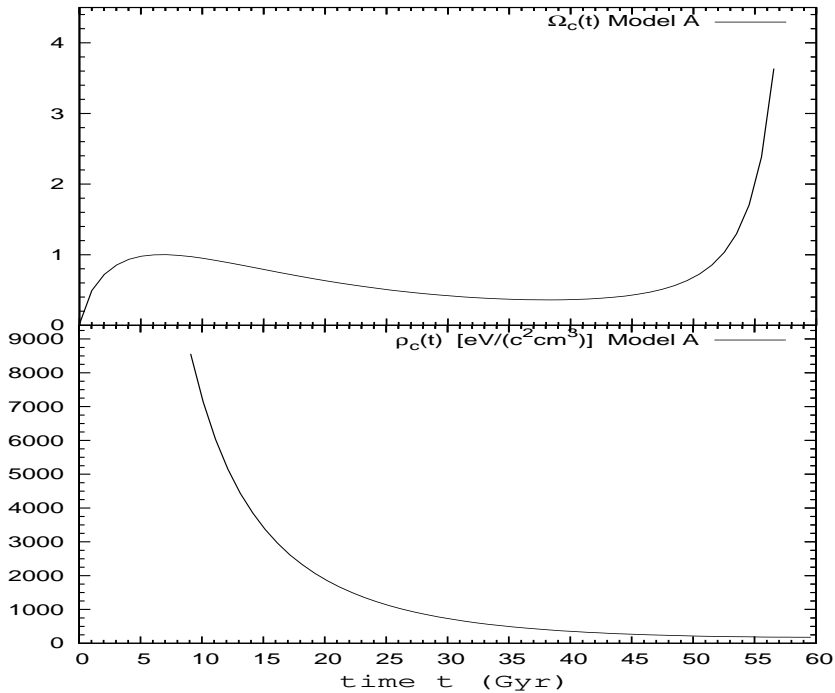}
\caption{Time evolution of the fractional spatial curvature density $\Omega_c(t)$ and the curvature density $\rho_c(t)$ in the Model A.}
\label{Figure 8}
\end{figure}

In the Model A we know the scale factor and the Hubble function in terms of the phase $\phi(t)=\Omega t$ so we can calculate the evolution of $\Omega_c$ over the entire period of the expansion from $\phi=0$ to $\phi=180^\circ$. In the Figure 8 we trace the evolution of $\Omega_c(t)$ over the time of the expansion. The Figure shows the local minima and maxima of $\Omega_c(t)$ corresponding to the acceleration-deceleration transition points. With $\Omega_{c,0}=0.831943$ at the present time $t_0=13.81$ Gyr we find $\rho_{c,0}=4030.91$ eVc$^{-2}$cm$^{-3}$. Assuming the entropic form of the Model A with $w_r=0$ this compares, with the same units, with $\rho_{0,0}=2564.45$ for Dark Energy, $\rho_{M,0}=2019.51$ for Dark Matter, $\rho_{m,0}=234.51$ for atomic matter, $\rho_{rad,0}=26.65$ for the radiation and with $\bar{\rho}_0=4845.17$ for the total energy density.

The Hubble function increases from $H=0$ at $t_\alpha=0$ to its maximum $H_{max}$=4.1294x10$^{53}$ Gyr$^{-1}$ at $t^*$=2.174859x10$^{-38}$ s at which time $\Omega_c(t^*)$=4.2747x10$^{-47}$. At z=1100 (t=0.000164) corresponding to the Cosmic Microwave Background the  Universe was still very flat with $\Omega_c(CMB)$=8.4250x10$^{-10}$. The early Universe is deeply flat and evolves into the anti-de Sitter spacetime at later times in the Model A. 

\section{Conclusions and Outlook}

We identify Dark Energy and Dark Matter with the Space. General theory of relativity asserts that Space is gravitational fields~\cite{einstein16}. The gravitational fields of Dark Energy and Dark Matter arise from a specific model of the quantum structure of the spacetime. This quantum structure of the spacetime is observable because it modifies the Hubble function relative to the $\Lambda$CDM Model. 

The model specifies by the relations (4.4)-(4.9) the entropic (dynamical) terms $\Sigma_0(z)$ and $\Sigma_M(z)$ of Dark Energy and Dark Matter in terms of the entropy $\Sigma_\lambda(z)$ of the average quantum state $\rho_\lambda(t)$ of Dark Energy. It predicts a new term in the Hubble function $(1+z)^{3(1+w_r)}\Omega_{r,0}>0$ with the equation of state $w_r=-\frac{1}{3}$ representing an internal galactic curvature. Consistency with the curvature term in the Robertson-Walker metric requires positive curvature density $\Omega_{c,0}$ the consequence of which are the constraints (1.3) and (1.4) on cosmological parameters $\Omega_{M,0}$ and $\Omega_{0,0}$.

The fits to Hubble data confirm the existence of the "residual" matter term with $w_r=-\frac{1}{3}$ and validate the constraints on the cosmological parameters. The fits to $d_A(z)$ determine positive values of $\Omega_{c,0}$ in in Models A, E1, E2 in agreement with the quantum model of the spacetime as well as $\Omega_{c,0} \approx 0$ for the $\Lambda$CDM Model. The fits used the Etherington relation (6.11) with $\eta_0=0.0150$. This value for $\eta_0$ is in excellent agreement with the very recent model independent test of the Etherington relation (6.11) which yields $\eta_0=0.0147^{+0.056}_{-0.066}$~\cite{ruan18}.

It is interesting to compare our results for $\Omega_{c,0}$ with a recent analysis~\cite{wang17}. These authors find that the best fit of $\Omega_{c,0}$ correlates with $H_0$ and the opacity $\epsilon$. For $H_0=73.24 \pm 1.74$ kms$^{-1}$Mpc and $\epsilon=-0.232\pm 0.075$ they find $\Omega_{c,0}=1.044\pm 0.514$ consistent with our result $\Omega_{c,0}=0.832\pm 0.240$ in the Model A with $H_0=67.81\pm 0.92$ kms$^{-1}$Mpc and $\epsilon=-0.015$. Within errors their result is also consistent with our $\Omega_{c,0}$ in the Model E1 and especially the Model E2.

The fits of the Model E to the analytical Model A (data $AH(z)$) of the Cyclic Universe developed in a related paper~\cite{svec17a} confirm the equivalence of the two models with an astonishing $\chi^2/dof=0.0000057$ in the range of the Hubble data $0\leq z \leq 2.34$. In the Section VII we examine the range of the applicability of the linear approximation used in the derivation of relations (4.6)-(4.9). Friedmann equations (7.1) allow us to determine the exact terms $\Sigma_0(z)$ and $\Sigma_M(z)$ in the Model A for any $z$ which we then compare with the calculations of these terms in the Model E2 extended beyond the limit of $z=2.34$. We find that the linear approximation works well for $z<20$ but rapidly breaks down for $z>20$. We identify the exact Model E with the Model A at all $z$ and all times $t$ and formulate the entropic form (1.1) for the Model A. We conclude that the Model A and the Model E represent the same Cyclic Universe with negative curvature and quantum structure of the spacetime.

In the Section VIII we show that the curvature parameter $R_0$ arises from the entropic terms $\Sigma_0$ and $\Sigma_M$ in the fits to the 
data $d_A(z)$ and therefore originates in the quantum structure of the spacetime. We assume that the comoving volume of our Universe is finite and given by $\mathcal{V}=R^3_0$. The proper volume $V(t)=a(t)^3\mathcal{V}=R(t)^3$ represents our observable Cyclic Universe. The spatial extent of the Universe in Model A is illustrated in the Table X and in a more detail in the Table IV of Ref.~\cite{svec17a}. The curvature parameter $R_0$ thus has a deep cosmological meaning: it connects the quantum structure of the spacetime with the "size" of the Cyclic Universe $V(t)$.

We propose that the infinite anti-de Sitter spacetime has a macroscopic lattice structure with each cell of comoving volume $\mathcal{V}=R^3_0$ corresponding to a Cyclic Universe within a Cyclic Multiverse. The curvature parameter $R_0$ thus determines also the structure of the Cyclic Multiverse.

In Section VIII we point out that the anti-de Sitter geometry leads to an increasingly larger violation of the Lorentz symmetry at increasing  scales $d > 100$ Mpc due to the finite value of $R_0=11.0187$ Gyr in the Model A. The violation of the Lorentz symmetry is observable at 1000Mpc ($z=0.33489$) and manifests itself in the violation of the Etherington relation. With $R_0$ arising from the Dark Energy and Dark Matter terms $\Sigma_0$ and $\Sigma_M$, the ultimate origin of the Lorentz symmetry violation is the quantum structure of the spacetime.

In general, the spatial curvatre density $\Omega_c(t)$ is not a static but a dynamic observable. We describe its time dependence in the Section IX and illustrate it with the Model A. The early Universe is deeply flat and evolves into the anti-de Sitter spacetime at later times in the Model A. There is no Standard Model inflation in the Model A. 

In $\Lambda$CDM Model the spatial curvature density is static and $\Omega_c(t)\equiv 0$ because there are no entropic terms $\Sigma_0$ and $\Sigma_M$ in this model. Since the curvature parameter $R_0=\infty$ there is no violation of the Lorentz symmetry at large scales in this Model. To explain the flat geometry at the early Universe the $\Lambda$CDM Model requires the assumption of the Inflation Model.

The observations to be made by the ongoing and upcoming astronomical surveys at high $z$ including Dark Energy Survey~\cite{DES}, Large Synoptic Survey Telescope~\cite{LSST}, Euclid Mission~\cite{Euclid}, Wide Field Infrared Survey Telescope~\cite{WFIRST,spergel15} and Square Kilometer Array~\cite{SKA} will significantly contribute to our understanding of Dark Energy and Dark Matter. Based on the fundamental theories of General relativity, Thermodynamics and Quantum information theory, and confirmed by the Hubble and angular diameter distance data, the cosmology of the Cyclic Universe with quantum structure of spacetime can be a useful participant in the analysis of these observations. 

\acknowledgements

I acknowledge with thanks the technical support of this research by Physics Department, McGill University.


\newpage
\appendix

\section{Table of the Angular diameter distance data $d_A(z)$, rescaled data $d_A(z,\eta_0)$ and data $Ad_A(z,\eta_0)$ predicted by the Model A}

\begin{table} 
\caption{Measured Angular diameter distance data $d_A(z)$, rescaled Angular diameter distance data $d_A(z,\eta_0)$ and the Angular diameter distance $Ad_A(z,\eta_0)$ predicted by the Model A in a fit to the rescaled data $d_A(z,\eta_0)$ assuming a correction factor $\eta_0=0.0150$. The parameters of the Model A are given by the best fit of the Model A to the Hubble data $H(z)$~\cite{svec17a}.}
\begin{tabular}{|c||c|c|c|c|c|c|c|}
\toprule 
$z$ & $d_A(z)$ & $\sigma_{d_A}(z)$ & $d_A(z,\eta_0)$ & $\sigma_{d_A}(z,\eta_0)$ & $Ad_A(z,\eta_0)$ & $A\sigma_{d_A}(z,\eta_0)$ & Reference \\
\colrule
    & ($Mpc$) & ($Mpc$) & ($Mpc$) & ($Mpc$) & ($Mpc$) & ($Mpc$) &    \\
\colrule  
0.023 & 103. & 42. & 103.0 & 42.0 & 98.60 & 0.002 & de Filippis {\sl et al.}~\cite{filippis05}\\
0.058 & 242. & 61. & 241.8 &  60.9 & 237.62 & 0.030 & -''- \\
0.072 & 165. & 45. & 164.8 &  45.0 & 289.80 & 0.056 & -''- \\
0.074 & 369. & 62. & 368.6 &  61.9 & 297.11 & 0.061 & -"- \\
0.084 & 749. & 385. & 748.1 & 384.5 & 333.08 & 0.088 & -"- \\
0.088 & 448. & 185. & 447.4 & 184.8 & 347.22 & 0.100 & -"- \\
0.091 & 335. &  70. & 334.5 &  69.9 & 357.73 & 0.110 & -"- \\
0.142 & 478. & 126. & 477.0 & 125.7 & 525.07 & 0.380 & -"- \\
0.171 & 809. & 263. & 806.9 & 262.3 & 611.49 & 0.630 & -"- \\
0.182 & 451. & 189. & 449.8 & 188.5 & 642.77 & 0.745 & -"- \\
0.183 & 604. &  84. & 602.3 &  83.8 & 651.16 & 0.778 & -"- \\
0.202 & 387. & 141. & 385.8 & 140.6 & 692.27 & 0.959 & -"- \\
0.202 & 806. & 163. & 803.5 & 162.5 & 703.00 & 1.010 & -"- \\
0.216 & 1465. & 407. & 1460.3 & 405.7 & 734.60 & 1.175 & -"- \\
0.224 & 1118. & 283. & 1114.3 & 282.1 & 755.21 & 1.294 & -"- \\
0.252 &  946. & 131. & 942.4 & 130.5 & 824.51 & 1.760 & -"- \\
0.282 & 1099. & 308. & 1094.4 & 306.7 & 894.21 & 2.354 & -"- \\
0.288 & 934.  & 331. & 930.0 & 329.6 & 907.62 & 2.485 & -"- \\
0.322 & 885.  & 207. & 880.7 & 206.0 & 980.48 & 3.305 & -"- \\
0.327 & 697.  & 183. & 693.6 & 182.1 & 990.77 & 3.437 & -"- \\
0.350 & 1048. &  58. & 1042.6 &  57.7 & 1032.83 & 4.024 & Chuang and Wang~\cite{chuang12}\\
0.350 & 1050. &  38. & 1044.5 &  37.8 & 1040.64 & 4.142 & Yu and Wang~\cite{yu16}\\
0.374 & 1231. & 441. & 1224.1 & 438.5 & 1082.49 & 4.826 & de Filippis {\sl et al.}~\cite{filippis05}\\
0.440 & 1205. & 114. & 1197.1 & 113.3 & 1197.66 & 7.259 & Yu and Wang~\cite{yu16}\\
0.451 & 1166. & 262. & 1158.2 & 260.2 & 1215.48 & 7.723 & de Filippis {\sl et al.}~\cite{filippis05}\\
0.546 & 1635. & 391. & 1621.7 & 387.8 & 1355.25 & 12.530 & de Filippis {\sl et al.}~\cite{filippis05}\\
0.550 & 1073. & 238. & 1064.2 & 236.1 & 1360.63 & 12.767 & de Filippis {\sl et al.}~\cite{filippis05}\\
0.570 & 1380. & 23.  & 1368.3 &  22.8 & 1384.42 & 13.870 & Yu and Wang~\cite{yu16}\\
0.570 & 1421. & 20.  & 1408.9 &  19.8 & 1389.61 & 14.123 & Anderson {\sl et al.}~\cite{anderson14}\\
0.600 & 1380. & 95.  & 1367.7 &  94.2 & 1424.99 & 15.991 & Yu and Wang~\cite{yu16}\\
0.730 & 1534. & 107. & 1517.4 & 105.8 & 1571.06 & 27.159 & Yu and Wang~\cite{yu16}\\
0.784 & 2479. & 1023. & 2450.2 & 1011.1 & 1624.94 & 33.296 & de Filippis {\sl et al.}~\cite{filippis05}\\
\botrule
\end{tabular}
\label{Table XI.}
\end{table}

\newpage
\section{Table of the parameters $F_k$, $k=1,5$ in the expansion of $d_L(z)$ for the fitted Models.}

\begin{table} 
\caption{Hubble parameter $H_0$ and the parameters $F_k(z=0)$, $k=1,5$ in the expansion (6.7)-(6.9) of $d_L(z)$ for the fitted Models.}
\begin{tabular}{|c||c|c|c|c|c|c|}
\toprule
Model & Model A & $\Lambda$CDM & E1 & E1 & E2 & E2 \\
\colrule
$w_r$ & NA & 0 & 0 & -$\frac{1}{3}$ & 0 & -$\frac{1}{3}$ \\
\colrule
$H_0$ & 67.810000 & 67.810000 & 67.810983 & 67.811017 & 67.810170 & 67.810203 \\
$F_1$ & 1.000000 & 1.000000 & 1.000000 & 1.000000 & 1.000000 & 1.000000 \\
$F_2$ & -0.708814 & -0.462630 & -0.578952 &  -0.578846 & -0.706866 & -0.706731 \\
$F_3$ & 0.793554 & -0.285700 & -0.248575 & -0.249699 & 0.760276 & 0.758884 \\
$F_4$ & -1.441459 & 1.442403 & 3.313933 & 3.319865 & -1.054001 & -1.045189 \\
$F_5$ & 5.007817 & -0.137028 & -8.550889 & -8.990080 & 1.222362 & 0.748794 \\
\botrule
\end{tabular}
\label{Table XII.}
\end{table}


\begin{thebibliography} {}

\bibitem{einstein16} A.~Einstein, {\sl Die Grundlage der allgemeinen Relativit\"{a}tstheorie}, Annalen der Physik, {\bf 49}, 1916. Translation {\sl The Foundation of the General Theoty of Relativity} in  "A Stubbornly Persistent Illusion: The Essential Works of Albert Einstein", Edited by S. Hawking, Running Press 2007.

\bibitem{svec17e} M.~Svec, {\sl Quantum Structure of Spacetime and its Entropy in a Cyclic Universe with Negative Curvature I: A Theoretical Framework}, arXiv:1810.06321 [gr-qc], (2018).

\bibitem{svec17a} M.~Svec, {\sl Serial Acceleration-Deceleration Transitions in a Cyclic Universe with Negative Curvature}, arXiv:1809.07178 [physics.gen-ph],  (2018).

\bibitem{weinberg08} S.~Weinberg, {\sl Cosmology}, Oxford University Press, 2008.

\bibitem{carroll04} S.M.~Carroll, {\sl Spacetime and Geometry}, Addison Wesley, 2004.

\bibitem{jimenez03} R.~Jimenez {\sl et al.}, {\sl Constraints on the equation of state of dark energy and the Hubble constant from stellar ages and the CMB}, ApJ {\bf 593}, 622 (2003).

\bibitem{simon05} J.~Simon, L.~Verde and R.~Jimenez, {\sl Constraints on the redshift dependence of the dark energy potential}, Phys.Rev.{\bf D71}, 123001 (2005).

\bibitem{gaztanaga09} E. Gazta\~{n}aga, A.~Cabr\'{e} and L.~Hui, {\sl Clustering of Luminous Red Galaxies IV: Baryon Acoustic Peak in the Line-of-Sight Direction and a Direct Measurement of $H(z)$},  Mon.Not.R.Astron.Soc. {\bf 399}, 1663 (2009).

\bibitem{stern10} D.~Stern {\sl et al.}, {\sl Cosmic Chronometers: Constraining the Equation of State of Dark Energy I: $H(z)$ Measurement},  J.Cosmol.Astropart.Phys. {\bf 2}, 008 (2010).

\bibitem{moresco12} M.~Moresco {\sl et al.}, {\sl Improved constraints on the expansion rate of the Universe to $z \sim 1.1$ from the spectroscopic evolution of cosmic chronometers}, J.Cosmol.Astropart.Phys. {\bf 8}, 006 (2012).

\bibitem{zhang14} Cong Zhang {\sl et al.}, {\sl Four new observational $H(z)$ data from luminous red galaxies in the Sloan Digital Sky Survey data release seven}, Research in Astronomy and Astrophysics 2014 {\bf 14}, 1221 (2014).

\bibitem{delubac15}  T.~Delubac {\sl et al.}, {\sl Baryon acoustic oscillations in the  Ly$\alpha$ forest of BOSS DR11 quasars}, Astronomy $\&$ Astrophysics  574, A59(2015).

\bibitem{moresco15} M.~Moresco, {\sl Raising the bar: new constraints on the Hubble parameter with cosmic chronometers at $z\sim 2$}, MNRAS, 450, L16 (2015). 

\bibitem{moresco16} M.~Moresco {\sl et al.}, {\sl A 6$\%$ measurement of the Hubble parameter at $z \sim 0.45$: direct evidence of the epoch of cosmic re-acceleration}, JCAP, 05, id.014  (2016).

\bibitem{planck15} Planck Collaboration, P.A.R.~Ade {\sl et al.}, {\sl Planck 2015 results, XIII. Cosmological parameters},Astronomy $\&$ Astrophysics 594, A13(2015), arXiv:1502.01589 [astro-ph.CO], (2015).

\bibitem{PDG2015} Particle Data Group, {\sl Review of Particle Properties}, Chinese Physics {\bf C40}, Number 10, p.120 (2016).

\bibitem{PDG2014} Particle Data Group, {\sl Review of Particle Properties}, Chinese Physics {\bf C38}, Number 9, p.111 (2014).

\bibitem{akaike74} H.~Akaike, IEEE Trans. Autom. Control {\bf 19}, 716  (1974).

\bibitem{schwartz78} G.~Schwartz, Annals of Statistics {\bf 6},461 (1978).

\bibitem{burnham02} K.P.~Burnham and D.R.~Anderson, {\sl Model selection and multimodel inference}, Springer Verlag, 2002.

\bibitem{etherington33} I.M.H.~Etherington, {\sl On the definition of distance in general relativity}, Phil.Mag. {\bf 15}, 761 (1933); reprinted in Gen.Relativ.Gravit. {\bf 35}, 1055 (2007).

\bibitem{ellis13} G.F.R.~Ellis, R.~Poltis, J.P.~Uzan and A.~Weltman, {\sl The  blackness of the cosmic microwave background spectrum as a probe of distance-duality relation}, Phys.Rev.{\bf 87}, 103530 (2013).

\bibitem{Lv16} Meng-Zhen Lv and Jun-Qing Xia, {\sl Constraints on Cosmic Distance Duality Relation from Cosmological Observations}, arXiv:1606.08102 [astro-ph.CO], (2016).

\bibitem{holanda16a} R.F.L.~Holanda and K.N.N.O.~Barros, {\sl Searching for cosmological signature of the Einstein equivalence principle breaking}, Phys.Rev.{\bf D94}, 023524 (2016).

\bibitem{holanda16b} R.F.L.~Holanda, V.C.~Busti, F.S.~Lima and J.S.~Alcaniz, {\sl Probing the distance-duality relation with high-z data}, JCAP {\bf 1709}, no.09, 039 (2017). 

\bibitem{yang17} Tao Yang, R.F.L.~Holanda and Bin Hu, {\sl Constraints on the cosmic distance duality relation with simulated data of gravitational waves from the Einstein Telescope}, arXiv:1710.10929 [astro-ph.CO], (2017).

\bibitem{wang17} Guo-Jian Wang, Jun-Jie Wei, Zheng-Xiang Li, Jun-Qing Xia, Zong-hong Zhu, {\sl Model-independent Constraints on Cosmic Curvature and Opacity}, Astrophys.J. 847(2017) no.1,45.

\bibitem{li17a} Xin Li, Li Tang and Hai-Nan Liu, {\sl Testing the anisotropy of the universe with the distance duality relation}, arXiv:1707.00390 [gr-qc], (2017).

\bibitem{li17b} Xin Li and Hai-Nan Lin, {\sl Testing the distance duality relation using type-Ia supernova and ultra-compact radio sources}, MNRAS 474, 313(2018). 

\bibitem{avgoustidis09} A.~Avgoustidis, L.~Verde and R.~Jimenez, {\sl  Consistency among distance measurements: transparency, BAO scale and accelerated expansion}, JCAP 0906, 012 (2009).

\bibitem{avgoustidis12} A.~Avgoustidis, G.~Luzzi, C.J.A.P.~Martins and A.M.R.V.L.~Monteiro, {\sl Constraints on the CMB temperature-redshift dependence from SZ and distance measurements}, JCAP {\bf 2}, 013 (2012).

\bibitem{filippis05} E.~de Filippis, M.~Sereno, M.W.~Bautz and G.~Longo, {\sl Measuring the Three-dimensional Structure of Galaxy Clusters. I. Application to a Sample of 25 Clusters}, ApJ, 625,108 (2005).

\bibitem{chuang12} Chia-Hsun Chuang and Yun Wang, {\sl Measurement of $H(z)$ and $D_A(z)$ from the Two-Dimensional Two-Point Correlation Function of Sloan Digital Sky Survey Luminous Red Galaxies}, MNRAS, 426, 226 (2012).

\bibitem{yu16} H.~Yu and F.Y.~Wang, {\sl New Model-independent Method to Test the Cosmic Curvature}, Astrophys.J. 828(2016) 85Y. 

\bibitem{anderson14} L.~Anderson {\sl et al.}, {\sl The clustering of galaxies in the SDSS-III Baryon Oscillation Spectroscopic Survey: Baryuon Acoustic Oscillations in the Data Release 10 and 11 Galaxy Samples}, MNRAS,441,24 (2014).

\bibitem{yu17} H.~Yu, B.~Ratra and F.Y.~Wang, {\sl Hubble parameter and acoustic oscillation measurement constraints on the Hubble constant, the deviation from the spatially flat $\Lambda$CDM  Model, the deceleration-acceleration transition redshift, and spatial curvature}, arXiv:1711.03437 [astro-ph.CO], (2017).

\bibitem{ruan18} Chen-Zong ~Ruan, Fulvio ~Melia and Tong-Jie ~Zhang, {\sl Model-independent Test of the Cosmic Distance Duality Relation}, arXiv:1808.09331 [astro-ph.CO], (2018).

\bibitem{cotaescu17} I.I.~Cot\u{a}escu, {\sl Anti-de Sitter
relativity}, arXiv:1706.01785 [gr-qc], (2017).

\bibitem{denissenya18} M.~Denissenya, E.V.~Linder and A.~Shafieloo, {\sl Cosmic Curvature Tested Directly from Observations}, JCAP 1803, 041(2018).

\bibitem{Qi18} Jing-Zhao Qi, Shuo Cao, Sixuan Zhang, Marek Biesiada, Yan Wu and Zong-Hong Zhu, {\sl A Revised Test of Cosmic Curvature at High Redshifts: The Distance Sum Rule}, arXiv:1803.01990 [astro-ph.CO], (2018).

\bibitem{DES} Dark Energy Survey (DES), http://www.darkenergysurvey.org.

\bibitem{LSST} Large Synoptic Survey Telescope (LSST), http://www.lsst.org.

\bibitem{Euclid} Euclid Mission, http://sci.esa.int/euclid/.

\bibitem{WFIRST} Wide Field Infrared Survey Telescope (WFIRST), wfirst.gafc.nasa.gov.

\bibitem{spergel15} D.~Spergel {\sl et al.}, {\sl Wide-Field InfraRed Survey Telescope-Astrophysics Focused Telescope Assets}, arXiv:1503.03757 [astro-ph.CO], (2015).

\bibitem{SKA} Square Kilometer Array (SKA), arXiv:1603:01951 [astro-ph.IM] (2016).


\end{thebibliography}
\end{document}